\documentclass[aps, prx,twocolumn,nofootinbib,floatfix,superscriptaddress]{revtex4-2}

\usepackage[utf8]{inputenc}

\usepackage{lineno}

\usepackage{graphicx}
\usepackage{indentfirst}
\usepackage{braket}
\usepackage{float}
\usepackage{amsmath}
\usepackage{physics}
\usepackage{amssymb}
\usepackage{CJK}
\usepackage{esint}
\usepackage{color}
\usepackage[T1]{fontenc}
\usepackage{subfigure}
\usepackage{amsfonts}
\usepackage{footmisc}
\usepackage{multirow}
\usepackage[outdir=./]{epstopdf}
\usepackage{svg}
\usepackage{algorithm}
\usepackage{algpseudocode}

\usepackage[english]{babel}
\usepackage{url}
\usepackage{comment}
\usepackage{array}
\usepackage{subfiles}
\usepackage{tikz}
\usetikzlibrary{shapes,arrows}
\usepackage{mathrsfs}
\usepackage[mathscr]{eucal}
\usepackage[hyperfootnotes=false]{hyperref}
\usepackage{stmaryrd}
\definecolor{darkblue}{rgb}{0,0,0.5}
\hypersetup{
colorlinks=true,
linkcolor=black,
filecolor=blue,
citecolor=darkblue,  
urlcolor=black,
}
\usepackage{mathtools}

\newtheorem{theorem}{Theorem}

\usepackage{trimclip}

\makeatletter
\DeclareRobustCommand{\shortto}{%
  \mathrel{\mathpalette\short@to\relax}%
}

\newcommand{\short@to}[2]{%
  \mkern2mu
  \clipbox{{.5\width} 0 0 0}{$\m@th#1\vphantom{+}{\shortrightarrow}$}%
  }
\makeatother

\usepackage{pict2e,picture,graphicx}

\makeatletter
\DeclareRobustCommand{\Arrow}[1][]{%
\check@mathfonts
\if\relax\detokenize{#1}\relax
\settowidth{\dimen@}{$\m@th\rightarrow$}%
\else
\setlength{\dimen@}{#1}%
\fi
\sbox\z@{\usefont{U}{lasy}{m}{n}\symbol{41}}%
\begin{picture}(\dimen@,\ht\z@)
\roundcap
\put(\dimexpr\dimen@-.7\wd\z@,0){\usebox\z@}
\put(0,\fontdimen22\textfont2){\line(1,0){\dimen@}}
\end{picture}%
}
\makeatother
\newcommand{\veryshortrightarrow}{\hspace{.2mm}\scalebox{.8}{\Arrow[.1cm]}\hspace{.2mm}}

\def\be{\begin{equation}}
\def\ee{\end{equation}}
\def\ba{\begin{eqnarray}}
\def\ea{\end{eqnarray}}
\def\bal{\begin{equation}\begin{aligned}}
\def\eal{\end{aligned}\end{equation}}
\def\o{\overline}

\def\bp{\begin{pmatrix}}
\def\ep{\end{pmatrix}}

\usepackage{bm}

\def\c2d{\rm C\veryshortrightarrow D}

\newtheorem{result}[theorem]{Result}

\newcommand{\calD}{{\cal D}}

\newcommand{\calI}{{\cal I}}
\newcommand{\calL}{{\cal L}}

\newcommand{\calJ}{{\cal J}}
\newcommand{\bbJ}{{\mathbb J}}
\newcommand{\calK}{{\cal K}}
\newcommand{\bbK}{{\mathbb K}}

\newcommand{\calR}{{\cal R}}
\newcommand{\calQ}{{\cal Q}}
\newcommand{\calS}{{\cal S}}

\newcommand{\1}{^{(1)}}

\newcommand{\QZ}[1]{{{\textcolor{black}{#1}}}}
\newcommand{\hw}[1]{{{\textcolor{black}{#1}}}}

\usepackage[normalem]{ulem}

\begin{document}

\begin{abstract}

The nature of dark matter is a fundamental puzzle in modern physics. A major approach of searching for dark matter relies on detecting feeble noise in microwave cavities. However, the quantum advantages of common quantum resources such as squeezing are intrinsically limited by the Rayleigh curse---a constant loss places a sensitivity upper bound on these quantum resources.
In this paper, we propose an in-situ protocol to mitigate such Rayleigh limit. The protocol consists of three steps: in-cavity quantum state preparation, axion accumulation with tunable time duration, and measurement. 
For the quantum source, we focus on the single-mode squeezed state (SMSS), and the entanglement-assisted case using signal-ancilla pairs in two-mode squeezed state (TMSS), where the ancilla does not interact with the axion. 
From quantum Fisher information rate evaluation, we derive the requirement of cavity quality factor, thermal noise level and squeezing gain for quantum advantage. When the squeezing gain becomes larger, the optimal axion accumulation time decreases to reduce loss and mitigate the Rayleigh curse---the quantum advantage keeps increasing with the squeezing gain.
Overall, we find that TMSS is more sensitive in the low temperature limit. In the case of SMSS, as large gain is required for advantage over vacuum, homodyne is sufficient to achieve optimality. For TMSS, anti-squeezing and photon counting is necessary to be optimal.  Thanks to the recent advance in magnetic-field-resilient in-cavity squeezing and rapidly coupling out for photon counting, the proposed protocol is compatible with axion detection scenario.






\end{abstract}

\title{
Quantum-enhanced dark matter detection with in-cavity control: mitigating the Rayleigh curse}
\author{Haowei Shi}
\affiliation{
Ming Hsieh Department of Electrical and Computer Engineering, University of Southern California, Los
Angeles, California 90089, USA}

\author{Anthony J. Brady}
\affiliation{
Ming Hsieh Department of Electrical and Computer Engineering, University of Southern California, Los
Angeles, California 90089, USA}

\author{Wojciech G\'{o}recki}
\affiliation{INFN Sezione di Pavia, Via Agostino Bassi 6, I-27100, Pavia, Italy}

\author{Lorenzo Maccone}
\affiliation{Dipartimento di Fisica, Universit\`a degli Studi di Pavia, Via Agostino Bassi 6, I-27100, Pavia, Italy}
\affiliation{INFN Sezione di Pavia, Via Agostino Bassi 6, I-27100, Pavia, Italy}

\author{Roberto Di Candia}
\affiliation{Department of Information and Communications Engineering, Aalto University, Espoo, 02150 Finland}
\affiliation{Dipartimento di Fisica, Universit\`a degli Studi di Pavia, Via Agostino Bassi 6, I-27100, Pavia, Italy}

\author{Quntao Zhuang}
\email{qzhuang@usc.edu}

\affiliation{
Ming Hsieh Department of Electrical and Computer Engineering, University of Southern California, Los
Angeles, California 90089, USA}
\affiliation{Department of Physics and Astronomy, University of Southern California, Los
Angeles, California 90089, USA}

\maketitle

\section{Introduction}

Microwave quantum engineering offers unprecedented quantum control~\cite{Blais2021cqedRMP,Casariego2023PropagMicrowaveRvw}, facilitating broad applications in quantum computing and quantum sensing. This capability enables high-fidelity state preparation, such as the creation of squeezed states, Fock states and Gottesman-Kitaev-Preskill state~\cite{gottesman2001,brady2024advances}, and non-trivial detection strategies, such as the counting of individual microwave photons. Microwave photon counting, which can be implemented \textit{in situ}~\cite{Blais2021cqedRMP} or for traveling photons~\cite{Kono2018FlyCount,Besse2018FlyCount,Lescanne2020CountNonrecip}, is crucial for quantum sensing and has gained recent prominence as a vital technology for precision measurements in high energy and particle physics~\cite{Ahmed2018QuSenseHEP, Bass2024NatRvw}.

A notable example where quantum techniques are expected to markedly impact fundamental physics~\cite{Ye2024Essay} is the search for new particles beyond the standard model, such as the axion, a promising dark matter (DM) candidate~\cite{Graham2015ExperSearchALPs,Bertone2018NewDMsearchEra,Sikivie2021RMP,Berlin2022SearchesSRF}. Photon counting has been considered for quantum-enhanced DM searches with cooled microwave cavity receivers~\cite{Lamoreaux2013LinAmpVsPhotonCount,zheng2016accelerating}. In axion DM searches, a cavity is immersed in a strong magnetic field (thus precluding the use of in-cavity \QZ{Josephson-junction-based} superconducting devices), inducing axion-to-photon conversion at an extremely feeble rate. The goal is to detect a faint excess of photons amidst the weak thermal background of the cooled cavity. 

Photon counting is advantageous as it allows for bypassing the vacuum noise inherent to linear detectors~\cite{Caves1982AmpNoise,Lamoreaux2013LinAmpVsPhotonCount,Beckey2023UserManQuLinDet}. Recently, photon counting was performed \textit{in situ} for a dark photon DM search by Dixit et al~\cite{dixit2021} and on travelling microwaves for an axion DM search by Braggio et al~\cite{Braggio2024QuAxionSearch}. The quantum-enhanced sensitivity is remarkable in these setups, with Fisher information enhancements over linear detection strategies scaling as $1/{N_{\rm T}}$, where ${N_{\rm T}}$ is the residual thermal background.\footnote{For reference, ${N_{\rm T}} \sim 10^{-4}$ for a 6 GHz cavity at 40 mK.} Nevertheless, the systems of Refs.~\cite{dixit2021,Braggio2024QuAxionSearch} operate as \textit{passive} receivers with measurement bandwidths that are ultimately limited by the linewidth of the microwave cavity, thus restricting the search speed for DM candidates. 

The ability to engineer quantum probes for \textit{active} sensing---i.e., stimulating the cavity---provides an extra degree of utility. Squeezing is known to improve the detection bandwidth for linear (homodyne) detectors without sacrificing sensitivity~\cite{zheng2016accelerating,malnou2019,backes2021,HAYSTAC2023PhaseII}; a concept that can be further extended to a quantum sensor network~\cite{brady2022entangled}. However, the performance of linear detectors is very limited even when enhanced with squeezing: it requires larger than 30 dB of squeezing for a homodyne scheme to match a passive scheme based on photon counting~\cite{shi2023ultimate}.
Fortunately, replacing homodyne with photon-counting, squeezing is also capable of improving the measurement bandwidth of photon-counting schemes without sacrificing sensitivity. Indeed, in the absence of loss, a squeezed state photon-counting receiver was proven optimal for noise sensing, when the phase-sensitive amplification is applied before photon-counting~\cite{gorecki2022,shi2023ultimate}. However, it was subsequently shown that the squeezed state receiver can be quite sensitive to cavity loss~\cite{shi2023ultimate,gardner2024StochWaveformEst}, and therefore limited by practical constraints. 



To mitigate the vulnerability to loss, 
Ref.~\cite{shi2023ultimate} proposed an entanglement-assisted receiver based on two-mode squeezing and photon counting. In that setup, a signal mode is entangled with an idler mode, the idler is stored in a separate high-Q cavity, and photon counting is implemented on the signal (see Fig.~\ref{fig:concept}(b)). With perfect idler storage, this configuration was shown to be quantum optimal, predicting a quantum-enhanced scan rate beyond all passive sensing configurations. However, there spectral photon counter (capable of counting the photons in each frequency bin) is assumed, which is beyond the capability of near-term devices. Moreover, when taking the idler storage cavity loss into consideration, the performance is still cursed by the Rayleigh limit---Ref.~\cite{gardner2024StochWaveformEst} shows that such Gaussian sources are all subject to a limit set by loss, lowering the performance below the diverging $1/{N_{\rm T}}$ advantage from passive receivers with photon counting when noise ${N_{\rm T}}$ is low. Therefore, the way towards large quantum advantage seems to rely on non-Gaussian state engineering, which generally requires \QZ{Josephson-junction-based} superconducting microwave cavities that are imcompatible with magnetic field in axion haloscopes~\cite{agrawal2024stimulated}.

In this paper, we propose an in-situ axion DM search protocol to mitigate the effect of Rayleigh curse~\cite{gardner2024StochWaveformEst} in DM detection, without relying on any non-Gaussian source. In particular, the proposed protocol completely circumvents the Rayleigh curse when the axion signal is fully incoherent. Our proposal prepares the quantum probe in cavity, accumulate the axion signal in cavity, and detects the accumulated signal, e.g., by rapidly coupling out the cavity field to a transmission line and counting itinerant photons, as shown in Fig.~\ref{fig:concept}. We formulate the transient cavity dynamics, which allows us to optimize the accumulation time, therefore going beyond a fixed quantum channel that is doomed by the Rayleigh curse~\cite{gardner2024StochWaveformEst}. The in-cavity probe preparation avoids deleterious injection losses from coupling propagating microwave quantum states into and out of the signal and idler cavities and enables tuning the overall loss by varying the accumulation time.

We find neat qualitative benchmarks for next-generation axion DM searches that are capable of leveraging a swath of quantum resources (including inter-cavity entangling operations, bandpass-limited photon counters, and high-Q cavities) in unison.
To quantify the advantage, we derive quantum precision limits of single-mode squeezed state (SMSS) source and the entanglement-assisted two-mode squeezed state (TMSS) source, which are easily accessible Gaussian-state sources~\cite{vahlbruch2016detection,eberle2013stable}, and show that they are achievable by the nulling receiver (anti-squeezing followed by photon counting) in the low temperature limit ${N_{\rm T}}\to 0$. The entanglement-assisted TMSS protocol enjoys a better robustness than the SMSS protocol. In the case of SMSS, as quantum advantage over vacuum requires large gain $G{N_{\rm T}}\gtrsim1$, homodyne can in fact already achieve close-to-optimal performance, not requiring photon counting. It is noteworthy that such robustness against loss rate circumvents (but does not violate) the Gaussian-state Rayleigh curse predicted in Ref.~\cite{gardner2024StochWaveformEst}, because the transient cavity dynamics here is no longer a fixed quantum channel in Ref.~\cite{gardner2024StochWaveformEst} and the overall loss here can be suppressed by optimizing waiting time. 
\QZ{Our protocol requires the generation of squeezing in large magnetic fields, which can be done using  Kinetic Inductance Parametric Amplifier (KIPA)~\cite{xu2023,frasca2024,vaartjes2024strong}.}
For practical application, we demonstrate significant advantages at feasible parameter setups for near-term devices.
In contrast, non-Gaussian state preparation typically requires \QZ{Josephson-junction-based} superconducting elements within the signal cavity, which are forbidden in an axion search due to the strong magnetic field.


\section{Overview}

\begin{figure*}
    \centering
    \includegraphics[width=\linewidth]{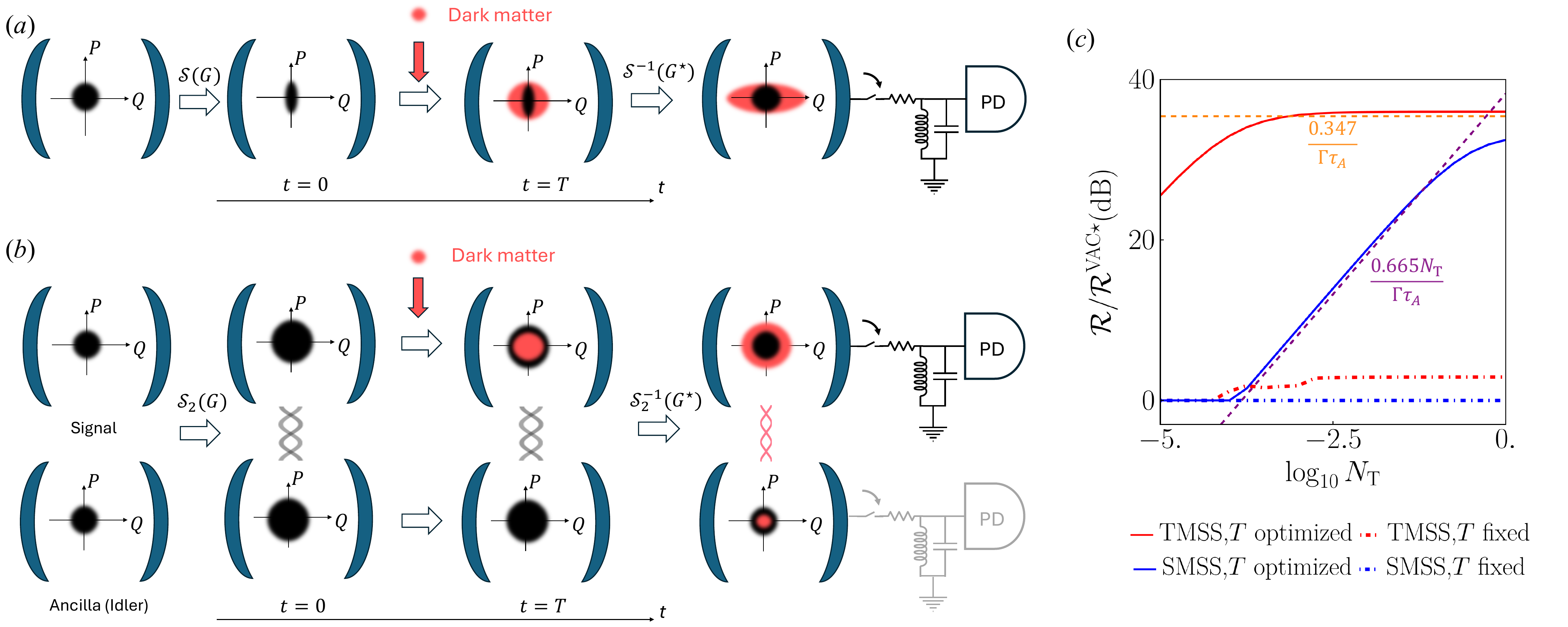}
    \caption{Conceptual schematic and performance advantage of the incavity protocol. (a) Squeezing-enhanced protocol using single-mode squeezed state (SMSS); (b) Entanglement-assisted protocol using two-mode squeezed state (TMSS). The quantum source states are generated from thermal background states of mean photon number ${N_{\rm T}}$ by single-mode squeezer $\calS(G)$ (or two-mode squeezer $\calS_2(G)$) of (quadrature) gain $G$. Then the dark matter of per-mode mean photon number $N_A$ couples into the cavity and accumulates for waiting time $T$, while the in-cavity mode suffers intrinsic loss at rate $\Gamma$ and gets mixed with an environment mode of thermal background photon number ${N_{\rm T}}$. After the accumulation, the optimal quantum measurement, which is known to be the nulling receiver~\cite{shi2023ultimate}, is made over the in-cavity mode, which composes first anti-squeezing $\calS^{-1}(G^\star)$ (or two-mode anti-squeezing $\calS_{2}^{-1}(G^\star)$) of optimized gain $G^\star$ and then rapidly coupling out for photon counting. The information carried in the ancilla is extremely weak thus the idler photon detector (colored in gray) is optional. PD: number-resolving photon detector. (c) Mitigation of Rayleigh curse at ${N_{\rm T}}\ll 1$. The quantum advantages over the vacuum-state $T$-optimized quantum Fisher information rate $\calR^{\rm VAC\star}$ of $T$-optimized cases (solid lines) are compared with the $T$-fixed cases (dot-dashed lines) where $T$ is fixed to vacuum optimum $T^{\rm VAC\star}$, for SMSS (blue) and TMSS (red) respectively. Asymptotic predictions Eqs.~\eqref{eq:SMSSadv_infG}\eqref{eq:TMSSadv_infG} assuming $G\to \infty, {N_{\rm T}}\ll 1, \Gamma\tau_A\ll 1$ are provided in dashed lines for SMSS (purple) and TMSS (orange). Squeezing gain $G$ is numerically optimized. Cavity parameters: $\Gamma\tau_A=10^{-4}, \Gamma_{\rm idler}\tau_A=10^{-8}$.}
    \label{fig:concept}
\end{figure*}

To detect axion DM, a microwave haloscope (a microwave cavity in a strong magnetic field~\cite{Sikivie2021RMP}) converts the DM field signals to feeble excess microwave power in the cavities. Therefore, DM search in such haloscopes becomes a noise sensing problem in the weak signal limit. In this regard, the Rayleigh limit curses the performance of DM search~\cite{tsang2023quantum,gardner2024StochWaveformEst} in the presence of loss: for any active detector with Gaussian source subject to loss on all modes, the sensitivity measured by quantum Fisher information (QFI) demonstrates a saturation to constant; For example, \hw{for a thermal loss channel with loss $1-\eta$ and additive noise $N_B$, an infinitely squeezed single-mode squeezed state (SMSS) generated from a thermal state of photon number $N_T$, has the QFI of noise parameter $N_B$ saturated to~\cite{GaussianQFI}
\be 
\calJ_{\rm SV}=\frac{2}{(1-\eta)^2(1+2{N_{\rm T}})^2},
\ee 
when $N_B=(1-\eta)N_{\rm T}$ is at the same temperature as the input,
which is limited to $2/(1-\eta)^2$} in the weak noise limit. From quantum Cramer-Rao bound, this limits the estimation precision of any additional noise---the means-square error $\delta^2\ge 1/\calJ_{\rm SV}$.
While for a passive detector with vacuum source, the performance in terms of QFI is divergent, $\calJ_{\rm VL}\sim1/{N_{\rm T}}\gg \calJ_{\rm SV} $, when the noise ${N_{\rm T}}$ is small. This casts a shadow on the promise of any quantum advantage.
One way out of the dilemma seems to be reducing loss: consider a short duration of detection such that the microwave cavities does not induce much loss. However, a shorter duration of detection also limits the signal accumulation from axion DM conversion and it is not clear if any quantum advantage can be achieved. In addition, the finite coherence time properties of axion DM becomes important in such a scenario and requires a systematic treatment. 

In this work, we perform a complete analyses of such a strategy and show that indeed by optimally tuning the accumulation time in a cavity, the Rayleigh curse can be largely mitigated. To consider a finite signal accumulation, we first refine the physical model of the dark matter field. We investigate the continuous-time model of random-phase dark matter field, where a phase jump happens probabilistically and yields a Lorentzian lineshape in the steady-state limit~\cite{dal2020resonators}, instead of simple additive noise model as in Refs.~\cite{shi2023ultimate,gardner2024StochWaveformEst}. Such a model of finite axion coherence time $\tau_A$ allows us to consider optimization of signal accumulation time rigorously. 


With the theoretical model refined, we consider possible improvement of the squeezing-based detection protocol.
As the coupling loss is the major obstacle that degrades the squeezing advantage, we consider the in-cavity protocol with the state generation conducted in cavity, instead of an input-output protocol that injects external quantum resources through a transmission line as proposed in Ref.~\cite{shi2023ultimate}. The conceptual schematic is shown in Fig.~\ref{fig:concept}. We consider two choices of quantum sources: (a) squeezing-enhanced protocol using SMSS and (b) entanglement-assisted protocol using TMSS. The entangled ancilla does not interact with the dark matter, but it benefits the detection by increasing the purity of the final state of the signal mode. In both cases, the quantum sources are generated in cavity, then interact with the dark matter field for an optimized accumulation (or waiting) time $T$. Finally, the cavity mode is processed by antisqueezing, then rapidly coupled out for detection. Such arrangement allows for axion search where the detection cavity is bathed in a strong magnetic field, for which we provide a detailed explanation at the end of this section.

In this paper we consider the occupancy number $N_A$ of the axion dark matter as the parameter of interest. In Sec.~\ref{sec:onresonance} we consider narrowband performance assuming the frequency of the dark matter signal is known. In Sec.~\ref{sec:scanrate}, we also discuss broadband performance assuming the frequency of the dark matter signal is \textit{a priori} unknown, and the signal must be searched for. As the total QFI $\calK(T)$ increases with the signal accumulation time $T$, in both cases, we consider the QFI rate $\calR\equiv \calK(T)/T$ as the figure of merit. The requirement of quantum advantage for the in-cavity detection protocol considered in this work can be formulated as the following.
\begin{result}
\label{main_result}
For any small but fixed thermal noise ${N_{\rm T}}$, fixed cavity loss rate $\Gamma$, the detection of the axion signal can be quantum enhanced over the vacuum limit by single-mode squeezed state (SMSS) source of quadrature squeezing gain $G$ when 
\begin{equation}
 \frac{\calQ_{\rm axion}}{{N_{\rm T}}} \lesssim\calQ_{\rm cav} ,\quad  G {N_{\rm T}} \gtrsim 1,
\label{eq:QReq_SMSSbeatVAC}
\end{equation} 
where $ \calQ_{\rm axion}\propto \tau_A$, $\calQ_{\rm cav}\propto 1/\Gamma$ are the quality factors of axion signal and the cavity.

For two-mode squeezed state (TMSS), of squeezing gain $G$ which is defined for quadrature variances analogous to the SMSS, the quantum advantage requirement is 
\begin{equation}
 \frac{\calQ_{\rm axion}}{{N_{\rm T}}}\lesssim\calQ_{\rm idler},\quad  G {N_{\rm T}} \gtrsim \frac{\calQ_{\rm cav}}{\calQ_{\rm idler}},
\label{eq:QReq_TMSSbeatVAC}
\end{equation} 
for lossy idler storage $\calQ_{\rm idler}/\calQ_{\rm cav}\ll 1/{N_{\rm T}} $ where $\calQ_{\rm idler}$ is the idler storage cavity quality factor. With ideal idler $\calQ_{\rm idler}/\calQ_{\rm cav}\to \infty$, the quantum advantage requirement reduces to 
\begin{equation}
 \calQ_{\rm axion} \lesssim\calQ_{\rm cav}, \quad G \gtrsim 1\,.
\label{eq:QReq_TMSSbeatVAC_idealidler}
\end{equation} 
\end{result}

In the limit of incoherent signal $\calQ_{\rm axion}\to 0$ (i.e., $\tau_A\to 0$) and large squeezing $G\gg1$, quantum advantage can always be achieved by tuning a proper signal accumulation time, therefore circumventing the Rayleigh limit. In fact, the advantage is infinite $\propto \calQ_{\rm cav}/\calQ_{\rm axion}\to \infty$ with unlimited $G\to \infty$. This is because smaller accumulation time reduces the overall loss $1-\eta(T)\simeq \Gamma T$ (at small $T\ll 1/\Gamma$), and indeed we find the optimal accumulation time $T^\star\to \tau_A\to 0$ at the limit of $G\to\infty$, $\tau_A\to 0$ as shown in Sec.~\ref{sec:formulas} later. For finite-value of parameters, the Rayleigh limit is mitigated, as we show in Fig.~\ref{fig:concept}(c). We plot the $G$-optimized quantum advantages over vacuum-state QFI rate using SMSS (blue solid) and TMSS (red solid) respectively. With a small signal coherence time $\Gamma\tau_A=10^{-4}$, by optimizing $T$ we achieve an advantage $\simeq 1/\Gamma\tau_A$. In contrast, with fixed $T$, i.e. fixed loss, we observe the Rayleigh curse (dashed lines): for any $G$ the squeezed-state QFI rate never surpasses the vacuum-state QFI rate for small ${N_{\rm T}}$. Here the TMSS demonstrates a much larger quantum advantage than SMSS, almost independent with ${N_{\rm T}}$, while the finite idler loss $\Gamma_{\rm idler}=10^{-4}\Gamma$ begins to limit such advantage as ${N_{\rm T}}$ decreases below $10^{-4}$.

In terms of experimental requirements, while cavity quantum electrodynamics (QED) techniques for state preparation requires \QZ{Josephson-junction-based} superconducting condition that is incompatible with magnetic field, 
recent experimental efforts have demonstrated magnetic-field-resilient in-cavity squeezing based on Kinetic Inductance Parametric Amplifier (KIPA)~\cite{xu2023,frasca2024,vaartjes2024strong}, which paves the way for in-cavity squeezing and antisqueezing in the presence of strong magnetic field. 
In comparison, non-Gaussian states such as Fock state~\cite{agrawal2024stimulated} or GKP states~\cite{gottesman2001,gardner2024StochWaveformEst} generally requires \QZ{Josephson-junction-based} superconducting cavity that is incompatible with magnetic field.
By rapidly coupling the signal out at the overcoupling limit, the photon counting can be implemented outside the cavity using Josephson photon-number amplifier~\cite{albert2024microwave} without loss of information. It is noteworthy to point out that in the case of SMSS, as quantum advantage over vacuum requires large gain $G{N_{\rm T}}\gtrsim1$ in Ineq.~\eqref{eq:QReq_SMSSbeatVAC}, homodyne can in fact already achieve close-to-optimal performance, not requiring photon counting.

In the TMSS protocol, the assumption of a good quantum memory for idler storage is plausible for an axion DM search because the signal cavity is typically copper (or more generally, non-superconducting), and the idler storage cavity---which need not be bathed in a strong magnetic field---can be superconducting. Indeed, recent progress on coupling two cavities, e.g. in Ref.~\cite{wurtz2021cavity,jiang2023}, demonstrates similar capabilities necessary for an entanglement-assisted scheme dedicated to axion DM searches.

Due to the advantage requiring $\calQ_{\rm cav}\gtrsim \calQ_{\rm axion}$, the scan-rate is approximately the same as the on-resonance axion signal case, and therefore all the results generalize to the scan-rate as well, as we detail in Sec.~\ref{sec:scanrate}.

\section{Formulation of in-cavity dynamics and detection}
\label{sec:incavityModel}


Our overall protocol involves preparing the cavity mode $A(0)$ in a certain quantum state, evolve for time $T$, and then perform measurement on the cavity field $A(T)$ to infer information about axion. To begin our analyses, we solve the dynamics for the intra-cavity field, $A(T)$. As $T$ is finite, previous models based on long-time analyses~\cite{brady2022entangled,shi2023ultimate} does not apply.

We consider a resonator in a bath with thermal population $N_{B}=(e^{\hbar\omega_c/k_B T_B}-1)^{-1}$, where $\omega_c$ is the resonator frequency and $T_B$ is the bath effective temperature. The resonator has a non-linear element able to operate in a large magnetic field. In addition, we assume that it is possible to generate high squeezing in a short time. This is the case, for instance, of KIPA~\cite{xu2023,frasca2024,vaartjes2024strong}.

The cavity mode $A(t)$ interacts with an axion field, governed by the Langevin equation
\begin{align}\label{model}
\partial_t A(t) = -\frac{\Gamma}{2}A(t) +\sqrt{\gamma_B}a_{B,in}(t)+\sqrt{\gamma_A}a_{A,in}(t)\,,
\end{align}
where $a_{B,in}(t)$ and $a_{A,in}(t)$ are respectively the bath field and the axion field, $\gamma_B,\gamma_A$ and $\Gamma=\gamma_A+\gamma_B$ are the coupling rates, and the equation is written in the rotating frame of frequency $\omega_c$. For a summary of notations and parameters please refer to Table~\ref{tab:dictionary}. We assume $a_{B,in}(t)
$ to be a white thermal noise, satisfying the commutation relation $[a_{B,in}(t),a_{B,in}^\dag(t')]=\delta(t-t')$ and having the autocorrelation $\expval{a_{B,in}^\dag(t')a_{B,in}(t)}={N_{\rm T}}\delta(t-t')$. The observable effect of the axion field is to displace the cavity mode with a random phase in time. This is modeled by choosing $a_{A,in}(t)$ in a coherent state with amplitude $\alpha(t)=|\alpha| e^{i\omega_A t + i\theta_A(t)}$, where $\omega_A$ is the detuning between the axion and the resonator frequencies. $|\alpha|^2$ gives the photon flux of axion field, which is determined by the axion mass and local DM density~\cite{Sikivie2021RMP}. \hw{In this paper we consider the estimation of the occupation number per axion mode, $N_A= |\alpha|^2\tau_A$, where $\tau_A$ is the axion coherence time. When the impinging direction of dark matter is unknown, the effective $|\alpha|^2$ in phase with the detector spatial mode can be a random variable, we define $N_A\equiv \expval{|\alpha|^2}\tau_A$ as the mean occupation number in general}. The phase $\theta_A(t)$ is a classical random variable. We heuristically model $\theta$ subject to a random phase jump as follows: If a jump has occurred at time $t$, the next jump will occur at time $t+t'$ with probability $p(t')=e^{-t'/\tau_A}/\tau_A$, where $\tau_A$ is the coherence time of the axion. At each jump, the phase is sampled uniformly at random in $[0,2\pi)$. Since the initial phase is also unknown and sampled uniform at random, the amplitude $\alpha$ is randomly distributed over the circle of radius $\alpha$ in the phase space. The state of $a_{A,in}$ is thereby fully dephased and diagonal in the photon-number basis, which is defined by the central moments. We derive the two-time correlation of $\alpha(t)$ as 
\begin{align}\label{alphattp}
\mathbb{E}[\alpha^*(\tau)\alpha(\tau')] &=  |\alpha|^2 e^{i\omega_A (\tau'-\tau)}e^{-|\tau'-\tau|/\tau_A}.
\end{align}

The solution to Eq.~\eqref{model} is the linear relation 
\begin{align}\label{sol}
\small A(T) \simeq &\, \sqrt{\eta(T)} A(0) + \sqrt{1-\eta(T)}a_B (T) \nonumber\\
\quad &+\sqrt{\frac{\gamma_A(1-\eta(T))}{\Gamma}} a_A(T)\,.
\end{align}
where $\eta(t)=e^{-\Gamma t}$. Here, we have assumed very weak coupling of the resonator with the axion field, i.e. $\gamma_A\ll\gamma_B$. Also, we have introduced the temporal-matched modes  $a_{(A,B)}(T)=\sqrt{\frac{\Gamma}{1-\eta(T)}}~\int_0^T \sqrt{\eta(T-\tau)}a_{(A,B),in}(\tau)d\tau$. The mode $a_B$ is in a thermal state with average photon number ${N_{\rm T}}$. The state of mode $a_A$ is fully dephased and defined by the central moments, due to the aforementioned random phase jump. The second moment of $a_A$ can be obtained via
\begin{align}\label{NA}
\langle a_A^\dag a_A\rangle & =\frac{\Gamma \eta(T)}{1-\eta(T)} \int_0^T\int_0^T e^{\Gamma(\tau+\tau')/2}\mathbb{E}[\alpha^*(\tau)\alpha(\tau')]d\tau d\tau'.
\end{align}
Computing \eqref{NA} using \eqref{alphattp} we obtain  
\begin{align}
\langle a_A^\dag a_A\rangle  = |\alpha|^2\tau_A \cdot g(T,\Gamma,\omega_A,\tau_A),
\label{eq:axionphoton_in-cavity}
\end{align}
where $g$ is a lengthy dimensionless expression (see Eq.~\eqref{eq:G_app} in Appendix~\ref{app:derive_QFI}). Finally, the effective axion occupation number mixed in the cavity mode $\hat A$ at time $T$ is 
\be 
n_A^{\rm eff}(T)\equiv \frac{\gamma_A(1-\eta(T))}{\Gamma} \langle a_A^\dag a_A\rangle\,.
\label{eq:nA_eff}
\ee
At the good cavity limit of $\Gamma\to 0$, we have 
\be 
n_A^{\rm eff}\to 2 \alpha ^2 \gamma _A \tau _A \left[\tau _A \left(e^{-\frac{T}{\tau _A}}-1\right)+T\right].
\label{eq:nA_eff_lowloss}
\ee 
We see that the photon flux $n_A^{\rm eff}/T$ is maximized at $T\to \infty$. Indeed, for $T\to 0$, $n_A^{\rm eff}\propto T^2$, while for $T\to \infty$, $n_A^{\rm eff}\propto T$. 

In principle, all central moments are needed to define the state of $a_A$. However, we note that the axion coupling is extremely weak such that $\sqrt{\gamma_A}|\alpha|\ll 1$, thus the moments of $\sqrt{\gamma_A}\alpha$ beyond second order are negligible. Thus we adopt the Gaussian approximation that ignores the higher order terms, which significantly simplifies the calculation workload as the resulting states are Gaussian and formulas of Gaussian-state QFIs are well known~\cite{gao2014bounds}. Specifically, we assume that the per-mode axion occupation number $|\alpha|^2\tau_A$ is a random variable subject to an exponential distribution with mean $N_A$. Indeed, we assume that ${\rm Pr}(|\alpha|^2\tau_A=x)=e^{-x/N_A}/N_A$, which means that $a_A$ is in a Gaussian thermal state with mean photon number $N_A g(T,\Gamma,\omega_A,\tau_A)$. 
To justify such approximation, in the Appendix~\ref{sec:justification} we show that the quantum advantage curves of the exact non-Gaussian model with $|\alpha|$ fixed are approximately the same as in the case of Gaussian in Ref.~\cite{shi2023ultimate}. With such approximation, Eq.~\eqref{sol} represents a bosonic thermal loss channel~\cite{weedbrook2012gaussian} from the initial field, $A(0)$, to the final field, $A(T)$.

\begin{table}[t]
    \renewcommand{\arraystretch}{1.6}
    \centering
    \begin{tabular}{c  c}
    \hline\hline
    Physical Parameters & Description  \\ \hline
    ${N_{\rm T}}$& Background thermal occupation\\
    $\gamma_B$& Bath coupling rate\\
    \hline 
    $\calQ_{\rm axion}\propto \tau_A$ & Axion quality factor \\
    $\tau_A$& Axion coherenc time\\
        \vspace{2mm}
    $\gamma_A$& Axion-cavity coupling rate\\
    $\omega_A$&  \shortstack{Axion center frequency \\ (detuned from cavity resonance)}\\
    $|\alpha|^2$& Axion flux\\
    $n_A^{\rm eff}$& \shortstack{Effective axion occupation\\ mixed in the cavity mode}\\
    $N_A=\tau_A\expval{|\alpha|^2}$& Average axion occupation\\
    \hline 
    $\omega_c$& Sensing cavity resonance frequency\\
    $\calQ_{\rm cav}\propto 1/\Gamma$& Sensing cavity quality factor\\
    $\Gamma=\gamma_A+\gamma_B$& Sensing cavity linewidth\\
    $\eta(t)=e^{-\Gamma t}$& transmissivity of cavity mode\\
    $\calQ_{\rm idler}$& Idler cavity quality factor\\
    $T$& Signal accumulation time \\
    \hline
    $\calK_{\rm VAC-HOM}$&QFI for vacuum homodyne\\
    $\calK_{\rm VAC}$&QFI vacuum limit\\
    $\calK_{\rm SMSS}$&QFI for SMSS\\
    $\calK_{\rm TMSS}$&QFI for TMSS\\
    $\calR_{\cdot}=\calK_{\cdot}/T$& Corresponding QFI rate\\
    \hline \hline
    Annihilation operator & Description\\
    \hline
    $A(t)$& Cavity mode\\
    $a_{B,in}(t)$ and $a_B$&  
    \shortstack{Bath field and \\ its temporal matched mode}\\
    $a_{A,in}(t)$ and $a_A$& \shortstack{Axion field and \\ its temporal matched mode}\\
    \end{tabular}
    \caption{Description of physical parameters and notations. QFI:quantum Fisher information. SMSS: single-mode squeezed state. TMSS: two-mode squeezed state.}
    \label{tab:dictionary}
\end{table}

\section{Quantifying measurement sensitivity}
\label{sec:formulas}
\label{sec:onresonance}

Our goal is to estimate the per-mode axion mean occupation number $N_A$, based on measurement on the final state $A(T)$. To enhance the performance, we propose to engineer the input state as SMSS and TMSS (joint with ancilla), which are compatible with the strong magnetic field~\cite{xu2023}. In this section, we analyze the performance of the protocols in terms of sensitivity; In Section~\ref{sec:scanrate}, we further address the scan rate. We begin with the evaluation of the QFI $\calK$ given SMSS and TMSS input, which provide an asymptotically achievable performance for all possible measurements on the final state, and then address the measurement scheme to achieve the bound. As axion signals are accumulated within a finite time $T$, we further convert the QFI results into the QFI rate $\calR\equiv \calK(T)/T$ as the figure of merit in choosing the optimal system setup. Furthermore, for each choice of quantum state, we optimize the waiting time $T$ and define the $T$-optimized QFI rate as $\calR^\star\equiv \max_T \calR(T)$, with the corresponding optimal waiting time $T^\star\equiv {\rm argmax}_T \calR(T)$. 

First, we provide analyses for the classical benchmark of passive detectors, where vacuum input \hw{(subject to thermal photon $N_T$)} and homodyne detection are adopted (resulting in the ``standard quantum limit'' for noise estimation~\cite{malnou2019}). For axion signal at detuning $\omega_A$ and waiting time $T$, we can solve the QFI $\calK_{\rm VAC-HOM}(\omega_A,T)$ analytically. However, as the full formula is lengthy (see Appendix~\ref{app:derive_QFI}), we present the simplified on-resonance ($\omega_A=0$) Fisher information
\bal 
&\calK_{\rm VAC-HOM}(\omega_A=0,T)= 32(\gamma_A \tau_A)^2\times \\
&\frac{e^{-2 \Gamma  T} \left(-2 \Gamma  \tau _A e^{\frac{1}{2} T
   \left(\Gamma -\frac{2}{\tau _A}\right)}+e^{\Gamma  T} \left(\Gamma  \tau
   _A-2\right)+\Gamma  \tau _A+2\right)^2}{\Gamma ^2 \tau _A^2 \left(2
   {N_{\rm T}}+1\right)^2 \left(\Gamma ^2 \tau_A^2-4\right)^2}.
\eal

As homodyne is not the optimal measurement for vacuum input, we also evaluate the QFI of the in-cavity protocol using vacuum state as $\calK_{\rm VAC}(\omega_A,T)$ in Appendix~\ref{app:derive_QFI}, which is known to be achievable by photon counting measurement~\cite{shi2023ultimate} (see Fig.~\ref{fig:measurement}). We use the vacuum-state QFI as a benchmark for the classical sources and demonstrate quantum advantages over it with squeezed sources. \hw{Given detuning $\omega_A$, the vacuum QFI is 
\be 
\calK_{\rm VAC}(\omega_A,T)=[\partial_{N_A} n_{A}^{\rm eff}(T)]^2 \frac{1}{{N_{\rm T}}(1+{N_{\rm T}})}
\ee
where $n_A^{\rm eff}(T)$ is the effective in-cavity axion photon number Eq.~\eqref{eq:nA_eff}.}
Plugging in Eq.~\eqref{eq:nA_eff}, the on-resonance vacuum QFI is 
\bal 
&\calK_{\rm VAC}(\omega_A=0,T)= 16(\gamma_A \tau_A)^2\times  \\
&\frac{ e^{-2 \Gamma T } \left(\tau_A  \Gamma-2 \tau_A  \Gamma e^{\frac{1}{2} T \left(\Gamma -\frac{2}{\tau_A }\right)}+\left(\Gamma\tau_A  -2\right) e^{\Gamma T }+2\right)^2}{\tau_A ^2 {N_{\rm T}}
   \left({N_{\rm T}}+1\right) \Gamma^2 \left(\Gamma^2\tau_A ^2 -4\right)^2}.
\eal
At the asymptotic limit of incoherent axion signal $\Gamma\tau_A\to 0$, we can show that the on-resonance vacuum-state QFI rate $\calR_{\rm VAC}(\omega_A=0,T)\equiv \calK_{\rm VAC}(\omega_A=0,T)/T$ is optimized at $T^\star_{\rm VAC}\simeq \left[-W_{-1}\left(-\frac{1}{2 \sqrt{e}}\right)-\frac{1}{2}\right] / \Gamma\approx 1.256/\Gamma$, where $W_{-1}(x)$ is the Lambert W function, which gives the -1th solution of $w$ for equation $x=we^w$. Such an asymptotic solution of $T^{\star}_{\rm VAC}$ can be understood as the following. Starting from time zero, the axion signal first coherent accumulates, and due to the coherence, one prefers to increase $T\gtrsim \tau_A$. Afterwards, axion signals can be modeled as incoherent additive noise with mean occupation number in Eq.~\eqref{eq:nA_eff_lowloss}, which increases with $T$ linearly, $\propto T N_A$ when $T\gtrsim \tau_A$. However, due to the weak coupling $\gamma_A\ll1$, $n_A^{\rm eff}\ll {N_{\rm T}}$, and the overall noise is still dominated by the thermal noise ${N_{\rm T}}$. Therefore, the QFI for estimating $N_A$ is $\propto T^2/{N_{\rm T}}({N_{\rm T}}+1)$, where $T^2$ comes from change of variable. This leads to a linear increase of QFI rate with $T$ until the cavity loss comes into play when $T\sim 1/\Gamma$ and stops increasing thereafter.


For the quantum squeezed sources, we denote the QFIs using SMSS and TMSS as $\calK_{\rm SMSS}(\omega_A,T)$, $\calK_{\rm TMSS}(\omega_A,T)$. We derive and put the formulas of $\calK_{\rm SMSS}(\omega_A,T)$ and $\calK_{\rm TMSS}(\omega_A,T)$ in Appendix~\ref{app:derive_QFI} as they are too lengthy. Below we provide asymptotic analyses and numerical evaluations to support Result~\ref{main_result}. 


\subsection{Asymptotic regimes of squeezing gain $G$}

We expect the performance to be enhanced by squeezing, while at low squeezing gain $G\to 1$, the squeezed state can perform even worse than the vacuum state in noise sensing since the loss destroys the purity of the squeezed states and induces extra noise, as predicted in Refs.~\cite{shi2023ultimate,gardner2024StochWaveformEst}. Thus we are interested in the break-even threshold $G_{\rm TH}$ where $\calR_{\rm SMSS/TMSS}^\star(G_{\rm TH})=\calR_{\rm VAC}^\star$ overcomes the optimal passive detector of vacuum photon counting. Prior results~\cite{shi2023ultimate} indicate that $G_{\rm TH}^{\rm SMSS}$ may be much larger than unity while $G_{\rm TH}^{\rm TMSS}$ may be close to unity. On the other hand, we expect the advantage from squeezing bounded at the limit of $G\to \infty$, due to imperfections of the system and finite axion coherence time. Thus we are also interested in the saturation point $G_{\rm SAT}$. In sum, there are two asymptotic regimes of interest for both SMSS and TMSS: 
\begin{enumerate}
    \item sufficiently large gain $G$ but not saturated, i.e. $G_{\rm TH}\ll G\ll G_{\rm SAT}$; 
    \item saturated gain $G\to \infty$. 
\end{enumerate}

\begin{figure}
    \centering
    \includegraphics[width=\linewidth]{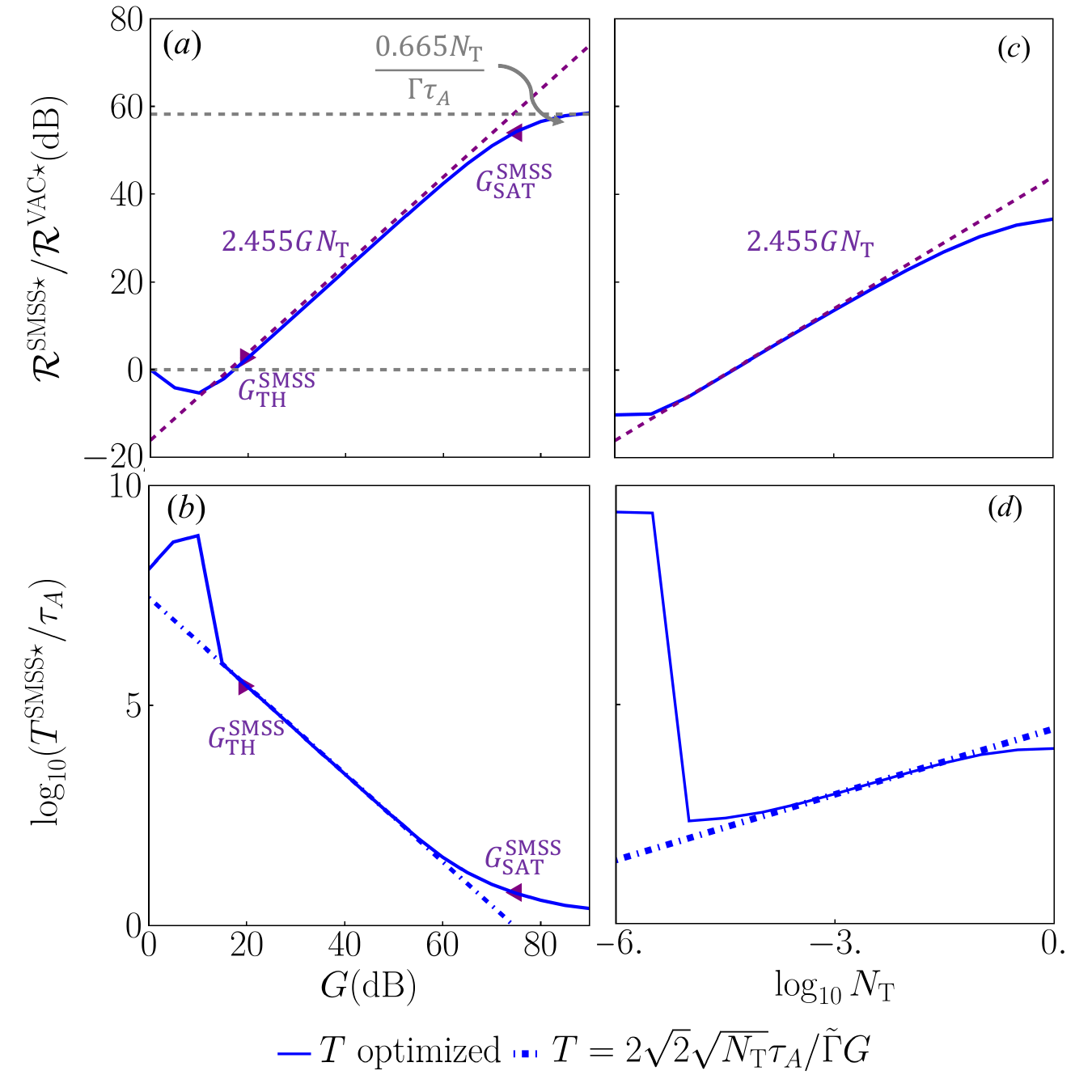}
    \caption{SMSS performance (a,b) versus squeezing gain $G$ with ${N_{\rm T}}=10^{-2}$ fixed; (c,d) versus thermal background photon number ${N_{\rm T}}$ with $G=40$dB fixed. (a,c) Advantage (in dB unit) of $T$-optimized quantum Fisher information rate $\calR^\star\equiv \max_T \calJ(T)/T$ over vacuum state input; (b,d) optimal waiting time $T^\star$ normalized by axion coherence time $\tau_A$ (in log10 scale). Solid lines: $T$ numerically optimized; Dot-dashed lines: asymptotic prediction with $T=2\sqrt{2{N_{\rm T}}}\tau_A/{\Gamma \tau_A}G$. Purple dashed lines: asymptotic prediction of linear advantage Eq.~\eqref{eq:SMSSadv_intermediateG} assuming non-saturated strong squeezing $G_{\rm SAT}\gg G\gg G_{\rm TH}, {N_{\rm T}}\ll 1, \Gamma\tau_A\ll 1$. In subplot (a,b) we mark $G_{\rm TH}$ ($\simeq 20$dB here) and $G_{\rm SAT}$ ($\simeq 55$dB here) by right-pointing triangle and left-pointing triangle respectively. Gray dashed lines: 0dB advantage and $G$-saturated advantage Eq.~\eqref{eq:SMSSadv_infG}. Cavity parameters: $\Gamma\tau_A=10^{-8}$. 
    }   
    \label{fig:advantage_SQZ}
\end{figure}

The optimal waiting time $T^\star(G)$ decreases as $G$ increases, because the quantum advantage is vulnerable to loss $1-\eta(T)\propto T$. For the same reason, $T^\star$ also depends on $\Gamma, {N_{\rm T}}$, while we focus on its dependence on $G$ here as $G$ is tunable. In the strong squeezing regime ($G\gtrsim G_{\rm SAT}$), the waiting time converges to the axion coherence time, $T^\star(G\to \infty)\to \tau_A^+$, because the axion coherence is concentrated within $\tau_A$ thus $T^\star\gtrsim \tau_A$ is required to sufficiently extract the axion coherence by coherent accumulation of duration $T^\star$ (see Appendix~\ref{sec:seqVspar}).


\subsection{Single-mode squeezing}

In Fig.~\ref{fig:advantage_SQZ}, we explore the on-resonance advantage of SMSS source over the vacuum under various squeezing gain, $G$, and background thermal noise, ${N_{\rm T}}$.  When $G\le G_{\rm TH}^{\rm SMSS}$ is small, we see the performance of SMSS is worse than vacuum limit, until the condition $G\gtrsim G_{\rm TH}^{\rm SMSS}$ is satisfied, as shown in subplot (a). After crossing the threshold, the advantage of SMSS grows linearly with gain $G$ till saturation at $G_{\rm SAT}^{\rm SMSS}$. We can obtain the analytic results of $G_{\rm TH}^{\rm SMSS}\simeq  1/{N_{\rm T}}$ and $G_{\rm SAT}^{\rm SMSS}\simeq 2\sqrt{2{N_{\rm T}}}/\Gamma\tau_A$ from asymptotics (see below), which agrees well with the numerical results as indicated by the two triangle points in Fig.~\ref{fig:advantage_SQZ}(a).

To understand the above behavior, we first perform asymptotic analyses to obtain the asymptotic optimal waiting time for $ G_{\rm TH}^{\rm SMSS} \ll G \ll G_{\rm SAT}^{\rm SMSS}$ where there is quantum advantage. Assuming the above asymptotic formula of $G_{\rm TH}^{\rm SMSS}$ and $G_{\rm SAT}^{\rm SMSS}$, we find that the on-resonance SMSS QFI rate $\calR_{\rm SMSS}(\omega_A=0,T)\equiv \calK_{\rm SMSS}(\omega_A=0,T)/T$ is optimized at $T^\star_{\rm SMSS}\simeq 2\sqrt{2 {N_{\rm T}}}/\Gamma G$, in the incoherent axion and low noise limits, $\Gamma\tau_A\to 0,{N_{\rm T}}\to 0$. The asymptotic optimal waiting time is verified in the numerical evaluations in Fig.~\ref{fig:advantage_SQZ} (b) and (d). Compared with the vacuum result, we see that larger squeezing requires a shorter optimal waiting time, as squeezing is more sensitive to loss. This decrease of optimal waiting time saturates towards $\tau_A$ when $G\gtrsim G_{\rm SAT}^{\rm SMSS}$. By solving $T^\star_{\rm SMSS}=\tau_A\simeq 2\sqrt{2 {N_{\rm T}}}/\Gamma G_{\rm SAT}^{\rm SMSS}$, we can obtain $G_{\rm SAT}^{\rm SMSS}\simeq 2\sqrt{2{N_{\rm T}}}/\Gamma\tau_A$, which is confirmed by the triangle point in Fig.~\ref{fig:advantage_SQZ}(a). On the other hand, larger thermal background increases the optimal waiting time, as thermal background makes the performance more tolerant to anti-squeezing noise. Note that the discontinuity in the waiting time is due to the competition between two local maximums, as we detail in Appendix~\ref{app:discontinuity}.

Under the asymptotic optimal waiting time, we can perform further asymptotics to confirm the numerical observations in Fig.~\ref{fig:advantage_SQZ}(a). First, when $G_{\rm SAT}^{\rm SMSS}\gg G\gg G_{\rm TH}^{\rm SMSS}$, we obtain the advantage over the vacuum state
\bal  
\!&\frac{\calR_{\rm SMSS}(\omega_A=0, T=2\sqrt{2 {N_{\rm T}}}/\Gamma G)}{\calR_{\rm VAC}(\omega_A=0,T=1/\Gamma)}\Big|_{ \substack{G_{\rm SAT}^{\rm SMSS}\gg G\gg G_{\rm TH}^{\rm SMSS}\\ \Gamma\tau_A, {N_{\rm T}}\to 0} }\\
&
\to 2.455 G {N_{\rm T}}\,,
\label{eq:SMSSadv_intermediateG}
\eal
which linearly grows with $G$. At the same time, Eq.~\eqref{eq:SMSSadv_intermediateG} also confirms the threshold of advantage $G_{\rm TH}^{\rm SMSS}\simeq  1/{N_{\rm T}}$. In Fig.~\ref{fig:advantage_SQZ}(c), we verify the scaling of advantage versus ${N_{\rm T}}$. Equation~\eqref{eq:SMSSadv_intermediateG} indicates that in order for SMSS to provide quantum advantage, we need $G{N_{\rm T}}\gtrsim 1$, which confirms the second part of Ineq.~\eqref{eq:QReq_SMSSbeatVAC} in Result~\ref{main_result}.


With $G\ge G_{\rm SAT}^{\rm SMSS}$ and optimal waiting time $T^\star=\tau_A$, the ultimate advantage of on-resonance SMSS QFI rate over the vacuum QFI rate can be well approximated by
\bal 
&\frac{\calR_{\rm SMSS}(\omega_A=0, T=\tau_A)}{\calR_{\rm VAC}(\omega_A=0,T=1/\Gamma)}\Big|_{G\to \infty;\Gamma\tau_A, {N_{\rm T}}\to 0} 
\\&= 
\frac{0.665 {N_{\rm T}}}{\Gamma  \tau _A}
   +O({N_{\rm T}}^2)\,.
\label{eq:SMSSadv_infG}
\eal
The result above then places a requirement for quantum advantage as ${N_{\rm T}}/\Gamma  \tau _A \gtrsim 1$. Considering the cavity quality factor $\calQ_{\rm cav}\propto1/\Gamma$ and axion quality factor, we obtain ${\calQ_{\rm axion}}/{{N_{\rm T}}} \lesssim\calQ_{\rm cav}$, which is the first part of Ineq.~\eqref{eq:QReq_SMSSbeatVAC} in Result~\ref{main_result}. Combining Eq.~\eqref{eq:SMSSadv_intermediateG}, we have obtained the condition of Ineq.~\eqref{eq:QReq_SMSSbeatVAC} in Result~\ref{main_result}.


\subsection{Two-mode squeezing}
\begin{figure}
    \centering
    \includegraphics[width=\linewidth]{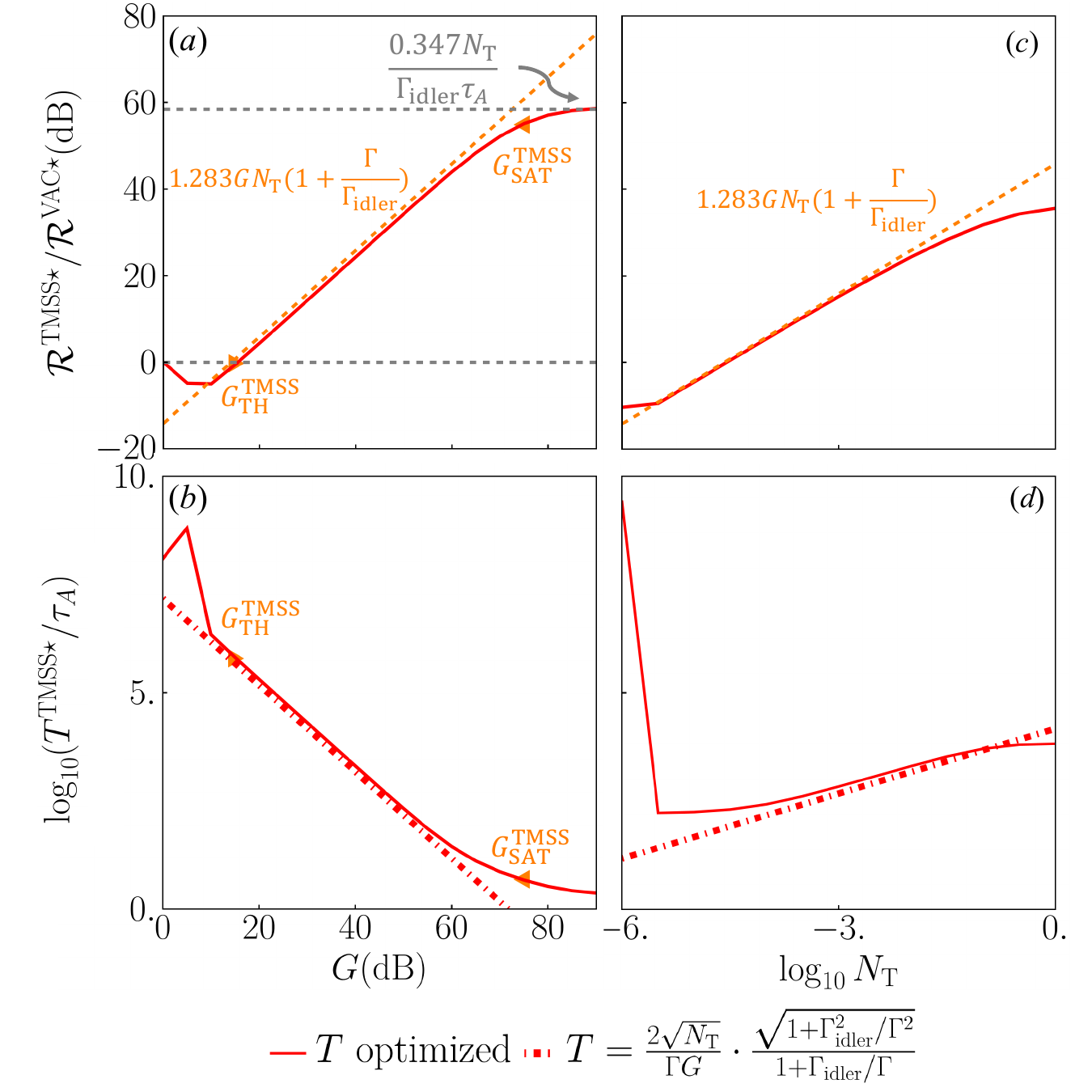}
    \caption{TMSS performance with lossy idler storage (a,b) versus squeezing gain $G$, with ${N_{\rm T}}=10^{-2}$ fixed; (c,d) versus thermal background photon number ${N_{\rm T}}$, with $G=40$dB fixed. (a,c) Advantage (in dB unit) of $T$-optimized quantum Fisher information rate $\calR^\star\equiv \max_T \calJ(T)/T$ over vacuum state input; (b,d) optimal waiting time $T^\star$ normalized by axion coherence time $\tau_A$ (in log10 scale). Dashed lines: asymptotic predictions Eq.~\eqref{eq:TMSSadv_intermediateG} assuming non-saturated strong squeezing $G_{\rm SAT}\gg G\gg G_{\rm TH}, {N_{\rm T}}\ll 1, \Gamma\tau_A\ll 1$ (Here $ 1.283 \left(1+\frac{\Gamma }{\Gamma _{\text{idler}} }\right)= 2.566$). Dotdashed lines: Asymptotic predictions on waiting time $T=\frac{2\sqrt{{N_{\rm T}}}}{\Gamma G}\cdot \frac{\sqrt{1+\Gamma_{\rm idler}^2/\Gamma^2}}{1+\Gamma_{\rm idler}/\Gamma}$. Cavity parameters: $\Gamma\tau_A=10^{-8}, \Gamma_{\rm idler}\tau_A=0.5\times 10^{-8}$.
    }   
    \label{fig:advantage_TMSS}
\end{figure}

\begin{figure}
    \centering
    \includegraphics[width=\linewidth]{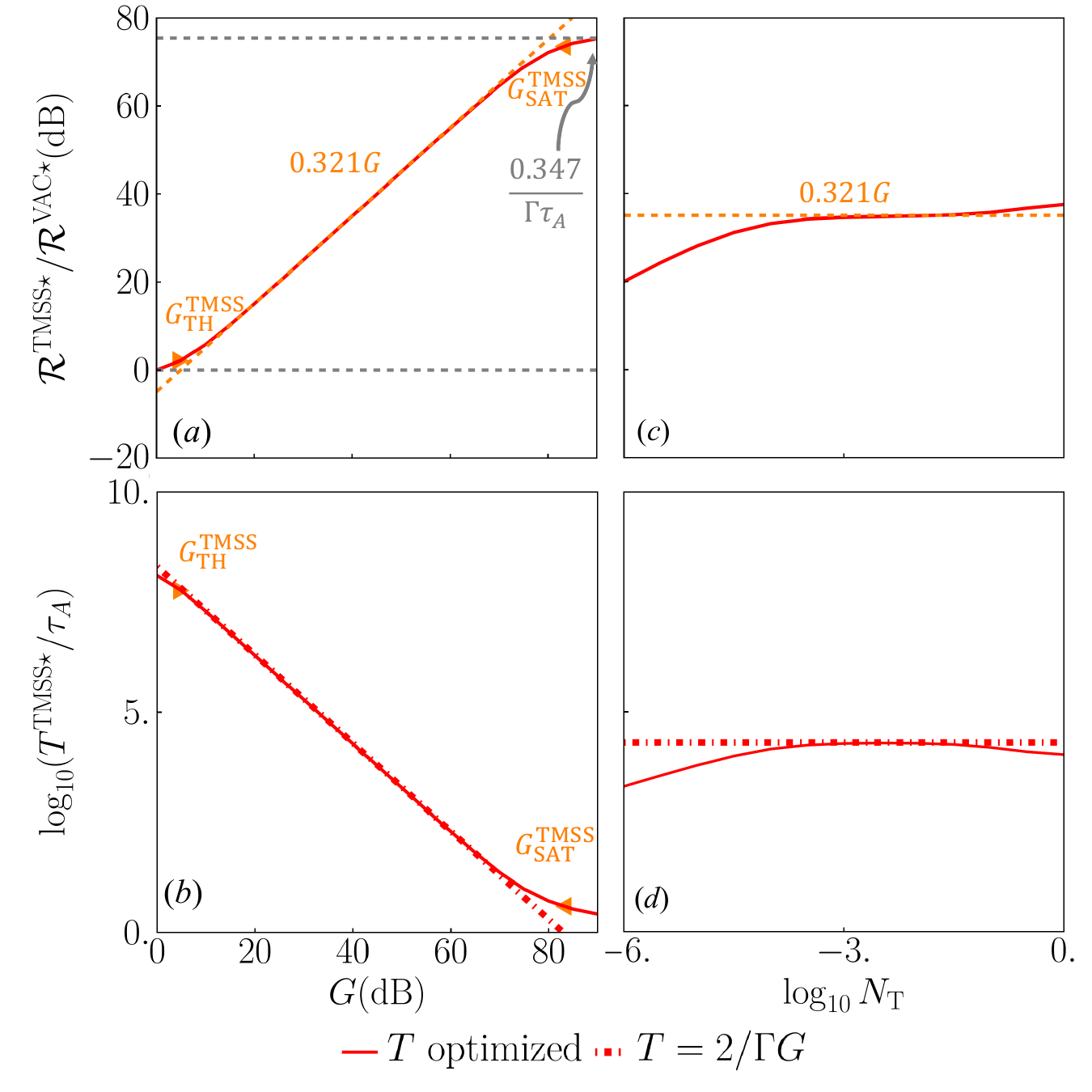}
    \caption{TMSS performance with ideal idler storage. The layout is identical to Fig.~\ref{fig:advantage_TMSS}. Here the asymptotic predictions of advantage (dashed lines) and $T$ (dot-dashed lines) are evaluated from Eq.~\eqref{eq:TMSSadv_intermediateG_idealidler} and $T=\frac{2}{\Gamma G}$ respectively. Cavity parameters: $\Gamma\tau_A=10^{-8}, \Gamma_{\rm idler}\tau_A=10^{-12}$.
    }   
    \label{fig:advantage_TMSS_idealidler}
\end{figure}

We first derive results for imperfect idler storage $\Gamma_{\rm idler}/\Gamma\gg {N_{\rm T}} $. For sufficiently large but not saturated $G$, i.e. $G_{\rm TH}^{\rm TMSS}\ll G\ll G_{\rm SAT}^{\rm TMSS}$, we find that the on-resonance TMSS QFI rate $\calR_{\rm TMSS}(\omega_A=0,T)$ is optimized at 
\be 
T^\star_{\rm TMSS}\simeq \frac{2 \sqrt{{N_{\rm T}} }}{ \Gamma  G} \frac{\sqrt{1+\Gamma _{\text{idler}}^2/\Gamma ^2}}{1+\Gamma _{\text{idler}}/\Gamma }
\ee 
when $\Gamma\tau_A\to 0$ and ${N_{\rm T}}\to 0$, as verified  in Figs.~\ref{fig:advantage_TMSS}(b)(d). Setting the waiting time saturating to the axion coherence time, $T^\star_{\rm TMSS}(G_{\rm SAT})=\tau_A$, we can solve the saturation gain 
\be 
G_{\rm SAT}^{\rm TMSS}\simeq \frac{2 \sqrt{{N_{\rm T}} }}{ \Gamma  \tau_A} \frac{\sqrt{1+\Gamma _{\text{idler}}^2/\Gamma ^2}}{1+\Gamma _{\text{idler}}/\Gamma }.
\ee 
Similar to SMSS, the TMSS advantage over vacuum state linearly grows with $G$ as 
\bal  
~&\frac{\calR_{\rm TMSS}(\omega_A=0, T=T^\star_{\rm TMSS})}{\calR_{\rm VAC}(\omega_A=0,T=1/\Gamma)}\Big|_{\substack{G_{\rm SAT}^{\rm TMSS}\gg G\gg G_{\rm TH}^{\rm TMSS}\\ \Gamma\tau_A,{N_{\rm T}}\to 0;~\Gamma_{\rm idler}/\Gamma\gg {N_{\rm T}}} }\\
&
\to 1.283 G {N_{\rm T}} \left(1+\frac{\Gamma }{\Gamma _{\text{idler}} }\right) \,.
\label{eq:TMSSadv_intermediateG}
\eal
We plot the predictions of linear growing advantage in dashed lines in Fig.~\ref{fig:advantage_TMSS}(a)(c), which agree with the numerical evaluation well within $G_{\rm TH}^{\rm TMSS}\ll G\ll G_{\rm SAT}^{\rm TMSS}$.
When $\Gamma=\Gamma_{\rm idler}$, as both cavities has the same loss, TMSS can be reduced to SMSS under balanced beamsplitter, and indeed we find the rate advantage for TMSS $\sim 2.566G{N_{\rm T}}$ in Eq.~\eqref{eq:TMSSadv_intermediateG}, which is almost equal to that of SMSS in Eq.~\eqref{eq:SMSSadv_intermediateG}. From equation $\calR_{\rm TMSS}(G=G_{\rm TH}^{\rm TMSS})=\calR_{\rm VAC}$ we solve the threshold gain $G_{\rm TH}^{\rm TMSS}\simeq [{N_{\rm T}}(1+\Gamma/\Gamma_{\rm idler})]^{-1}$.
When $G$ approaches the saturation gain $G_{\rm SAT}^{\rm TMSS}$, the QFI rate is no longer represented by Eq.~\eqref{eq:TMSSadv_intermediateG}. Instead, asymptotic analyses at $G\to \infty$ leads to 
\be 
\frac{\calR_{\rm TMSS}(\omega_A=0, T=\tau_A)}{\calR_{\rm VAC}(\omega_A=0,T)}\simeq 
\frac{0.347 {N_{\rm T}} }{ \Gamma _{\text{idler}} \tau _A}.
\ee
We plot the saturated advantage in gray dashed line in Fig.~\ref{fig:advantage_TMSS}(a), which agrees with the numerical evaluations at large $G$.
The results above enforce constraints for TMSS to be advantageous as ${N_{\rm T}}/\Gamma_{\rm idler}  \tau _A \gtrsim 1$. Note that $\calQ_{\rm cav}\propto1/\Gamma$, $\calQ_{\rm axion}\propto \tau_A$, we obtain ${\calQ_{\rm axion}}/{{N_{\rm T}}} \lesssim\calQ_{\rm idler}$, which is the first part of Ineq.~\eqref{eq:QReq_TMSSbeatVAC} in Result~\ref{main_result}. Combining it with Eq.~\eqref{eq:TMSSadv_intermediateG}, we obtain the condition of Ineq.~\eqref{eq:QReq_TMSSbeatVAC} in Result~\ref{main_result}.

For ideal idler storage $\Gamma_{\rm idler}/\Gamma\ll {N_{\rm T}}$, the optimal waiting time is $T^\star_{\rm TMSS}\simeq 2/G\Gamma$, the advantage is 
\bal 
~&\calR_{\rm TMSS}(\omega_A=0, T=\tau_A)|_{\substack{G_{\rm SAT}^{\rm TMSS}\gg G\gg G_{\rm TH}^{\rm TMSS}\\ \Gamma\tau_A,\Gamma_{\rm idler}/\Gamma,{N_{\rm T}}\to 0}} \\
&\to 0.321 G +O({N_{\rm T}})\,.
\label{eq:TMSSadv_intermediateG_idealidler}
\eal
In this case, both the optimal waiting time and the advantage do not depend on ${N_{\rm T}}$ at all. By setting $\calR_{\rm TMSS}(G=G_{\rm TH})=\calR_{\rm VAC}$ and $T^\star_{\rm TMSS}(G_{\rm SAT})=\tau_A$, we solve the threshold gain $G_{\rm TH}\simeq 1/0.321$ and the saturation gain $G_{\rm SAT}\simeq 2/\Gamma\tau_A$.
This explains the saturation at $G\sim 80$dB in Fig.~\ref{fig:advantage_TMSS_idealidler}(a). The saturated advantage is independent on ${N_{\rm T}}$:
\be 
\frac{\calR_{\rm TMSS}(\omega_A=0, T=\tau_A)}{\calR_{\rm VAC}(\omega_A=0,T)}|_{G\to \infty;\Gamma\tau_A,\Gamma_{\rm idler}/\Gamma,{N_{\rm T}}\to 0}\to \frac{0.347}{\Gamma  \tau _A}\,.
 \label{eq:TMSSadv_infG}
\ee
The results above enforce constraints for TMSS to be advantageous as ${N_{\rm T}}/\Gamma_{\rm idler}  \tau _A \gtrsim 1$. Note that $\calQ_{\rm cav}\propto1/\Gamma$, $\calQ_{\rm axion}\propto \tau_A$, we obtain ${\calQ_{\rm axion}}/{{N_{\rm T}}} \lesssim\calQ_{\rm idler}$, which is the first part of Ineq.~\eqref{eq:QReq_TMSSbeatVAC} in Result~\ref{main_result}. Combining it with Eq.~\eqref{eq:TMSSadv_intermediateG}, we obtain the condition of Ineq.~\eqref{eq:QReq_TMSSbeatVAC} in Result~\ref{main_result}.


\begin{figure}
    \centering \includegraphics[width=\linewidth]{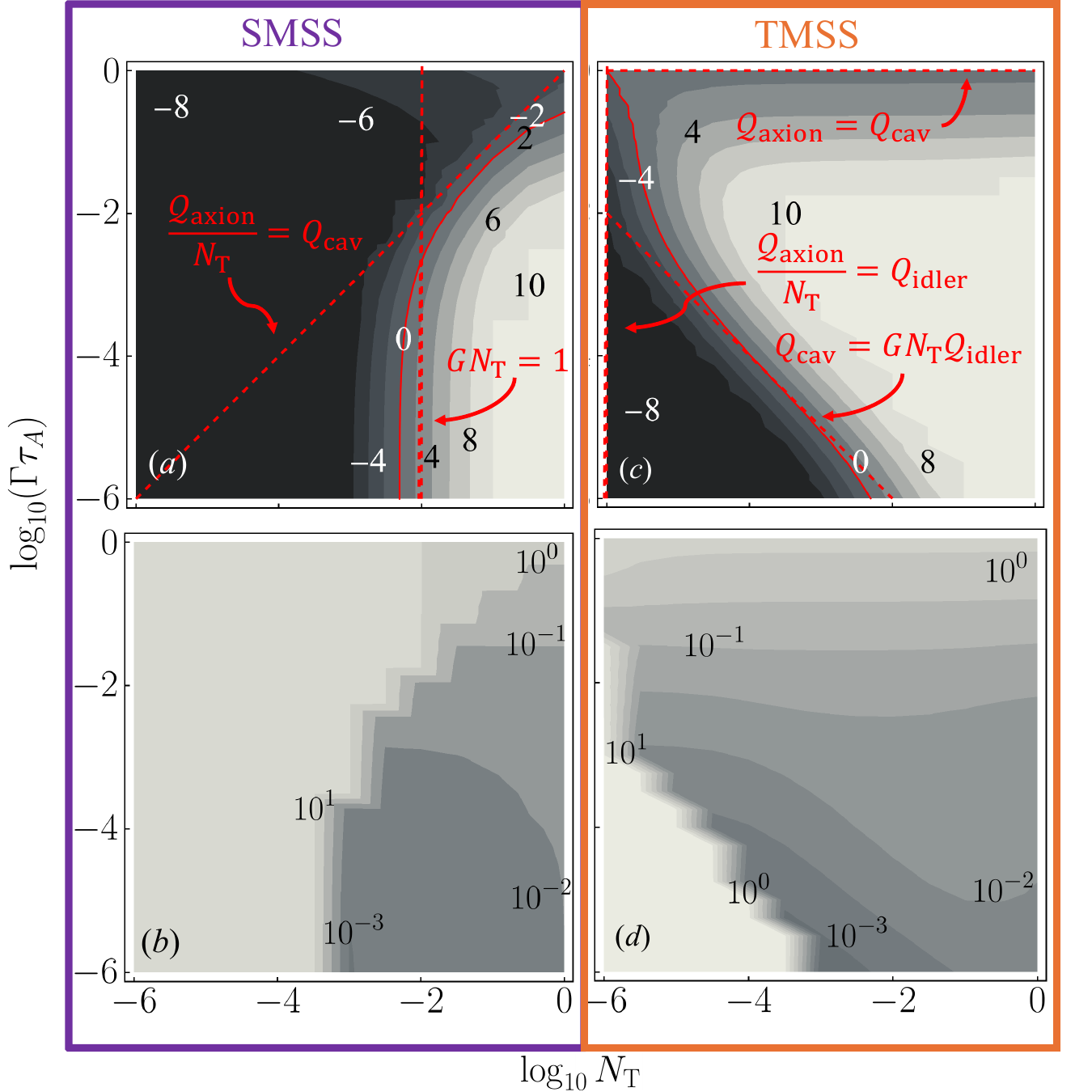}
    \caption{(a-b) SMSS performance ntage in comparison with the asymptotic quantum advantage requirement Result~\ref{main_result}. (a) Quantum advantage (in dB unit , x (dB)= 10 $\log_{10} x$) of $T$-optimized quantum Fisher information rate $\calR^\star$ over vacuum state input versus various cavity linewidth $ \Gamma$ (normalized by axion linewidth $1/\tau_A$) and thermal background noise ${N_{\rm T}}$. The break-even thresholds of 0dB quantum advantage are marked in red solid curves, while the asymptotic predictions on them in Ineq.~\eqref{eq:QReq_SMSSbeatVAC} are indicated in red dashed curves. (b) Optimal normalized waiting time $T^\star/(1/\Gamma)$. (c-d) TMSS performance in comparison with Result~\ref{main_result}, with similar layout with subplot (a-b), where the asymptotic predictions on break-even thresholds are from Ineqs.~\eqref{eq:QReq_TMSSbeatVAC}\eqref{eq:QReq_TMSSbeatVAC_idealidler}. For all four subplots $G=20$dB. For TMSS, ${\Gamma}_{idler}\tau_A= 10^{-6}$. 
    }    \label{fig:advantage_vs_Nb_z}
\end{figure}

\subsection{Result~\ref{main_result} and mitigating the Rayleigh curse}

While our asymptotic analyses have confirmed Result~\ref{main_result}, here we provide a direct contour plot to show case how these conditions limit the region of quantum advantage for resonant detection, i.e. $\omega_A=0$.

In Fig.~\ref{fig:advantage_vs_Nb_z}, we plot contours of advantage versus the dimensionless loss rate $\Gamma\tau_A$ and pre-squeezing thermal background ${N_{\rm T}}$ for $G=20$dB. In subplot (a) we consider the SMSS source, the advantage is limited to the good cavity regime ${\Gamma \tau_A}\ll 1$, and the advantage significantly degrades as the environment thermal noise decreases ${N_{\rm T}}\to 0$. Indeed, we can see the conditions in Ineqs.~\eqref{eq:QReq_SMSSbeatVAC} (red dashed) precisely captures the region of quantum advantage (red solid). 
By contrast, in subplot (c), the TMSS source shows advantage extending to noiseless case ${N_{\rm T}}\ll 1$, which is the regime of our interest in typical microwave haloscopes cooled to low temperature. At the bottom left corner in subplot (c), we observe a rapid degradation of advantage, this is because the idler loss rate ${\Gamma}_{\rm idler}\tau_A=10^{-6}$ begins to be comparable to the signal loss rate ${\Gamma \tau_A}$ at this disadvantageous corner. To keep the advantage, we see that ${\Gamma}_{\rm idler}/{\Gamma}\lesssim G{N_{\rm T}}$ is required. Indeed, the boundary of quantum advantage (red solid) is well captured by the Ineqs.~\eqref{eq:QReq_TMSSbeatVAC} as indicated by the red dashed curves. The additional condition in Ineq.~\eqref{eq:QReq_TMSSbeatVAC_idealidler} also shows up on the top of subplot (c) as in this region the idler storage cavity is close to ideal, compared to the sensing cavity. The optimal waiting time for both SMSS and TMSS are shown in subplots (b) and (d) correspondingly.

\begin{figure}[t]
    \centering    
    \includegraphics[width=0.8\linewidth]{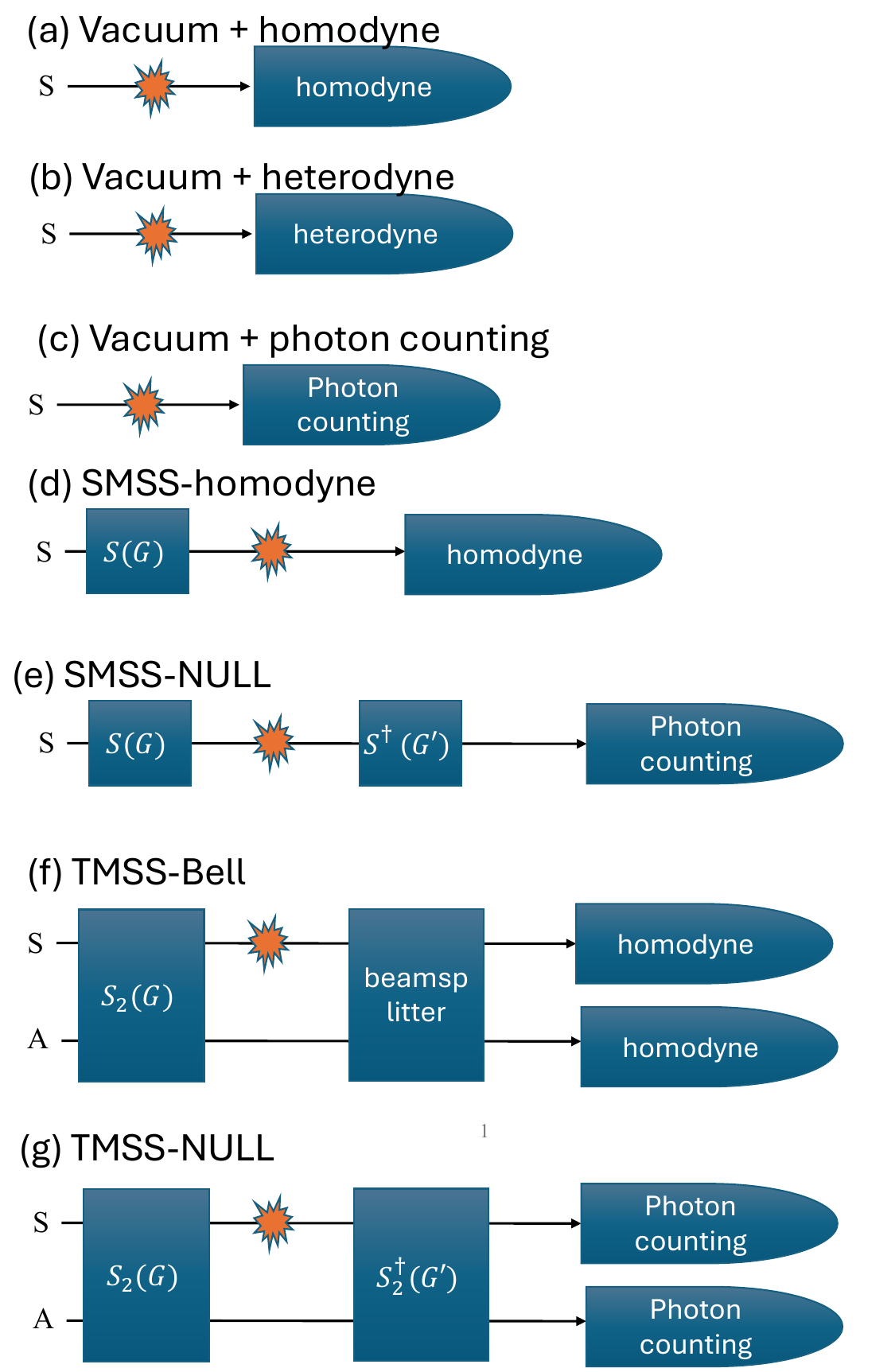}
    \caption{Schematic of the measurement protocols being considered. All modes $S$ and $A$ are initially in vacuum (up to inevitable thermal noise due to limited cooling).
    }
    \label{fig:measurement}
\end{figure}

Remarkably, in the limit of incoherent axion $\tau_A\to 0$, the SMSS advantage $\propto G\sqrt{{N_{\rm T}}}$ in Eq.~\eqref{eq:SMSSadv_intermediateG} circumvents the Rayleigh curse in Ref.~\cite{gardner2024StochWaveformEst}, as we have stated in Result~\ref{main_result}. The reason is that here the overall cavity loss is $1-\eta(T)=1-e^{-\Gamma T}$, which is always negligible for any finite cavity loss rate $\Gamma$ at the large gain $G$ limit such that $T\propto 1/G\to 0$. This conclusion also holds for TMSS with the advantage in Eq.~\eqref{eq:TMSSadv_intermediateG} not even affected by noise ${N_{\rm T}}$. For strong squeezing limit $G\to\infty$, similarly both SMSS and TMSS advantages in Eqs.~\eqref{eq:SMSSadv_infG}\eqref{eq:TMSSadv_infG} can be significant even at ${N_{\rm T}}\ll 1$ with $\Gamma\tau_A\to 0$.

\subsection{Measurement designs}

With the input states and QFI rates in hand, we now proceed to analyze the measurement protocols. As shown in Fig.~\ref{fig:measurement}, we consider the linear homodyne measurement and the photon number counting measurement for the three inputs: vacuum, SMSS, and TMSS respectively. For vacuum, we also consider heterodyne measurement as it has been found to yield 3dB advantage over homodyne when the environment is noisy. For SMSS and TMSS, we apply antisqueezing $S^\dagger$, $S_2^\dagger$ respectively before photon counting to extract the quantum advantage from quantum coherence, which is known as nulling receiver~\cite{shi2023ultimate}. As a benchmark, we also consider homodyne measurement for SMSS and the Bell measurement (a balanced beamsplitter followed by two homodyne detectors) for the TMSS input~\cite{pirandola2011quantum,shi2020entanglement}
Similarly, for each specific measurement, we adopt the same notation to denote the achievable classical Fisher information as $\calK$ and the Fisher information rate $\calR$.

\begin{figure}[t]
    \centering    \includegraphics[width=\linewidth]{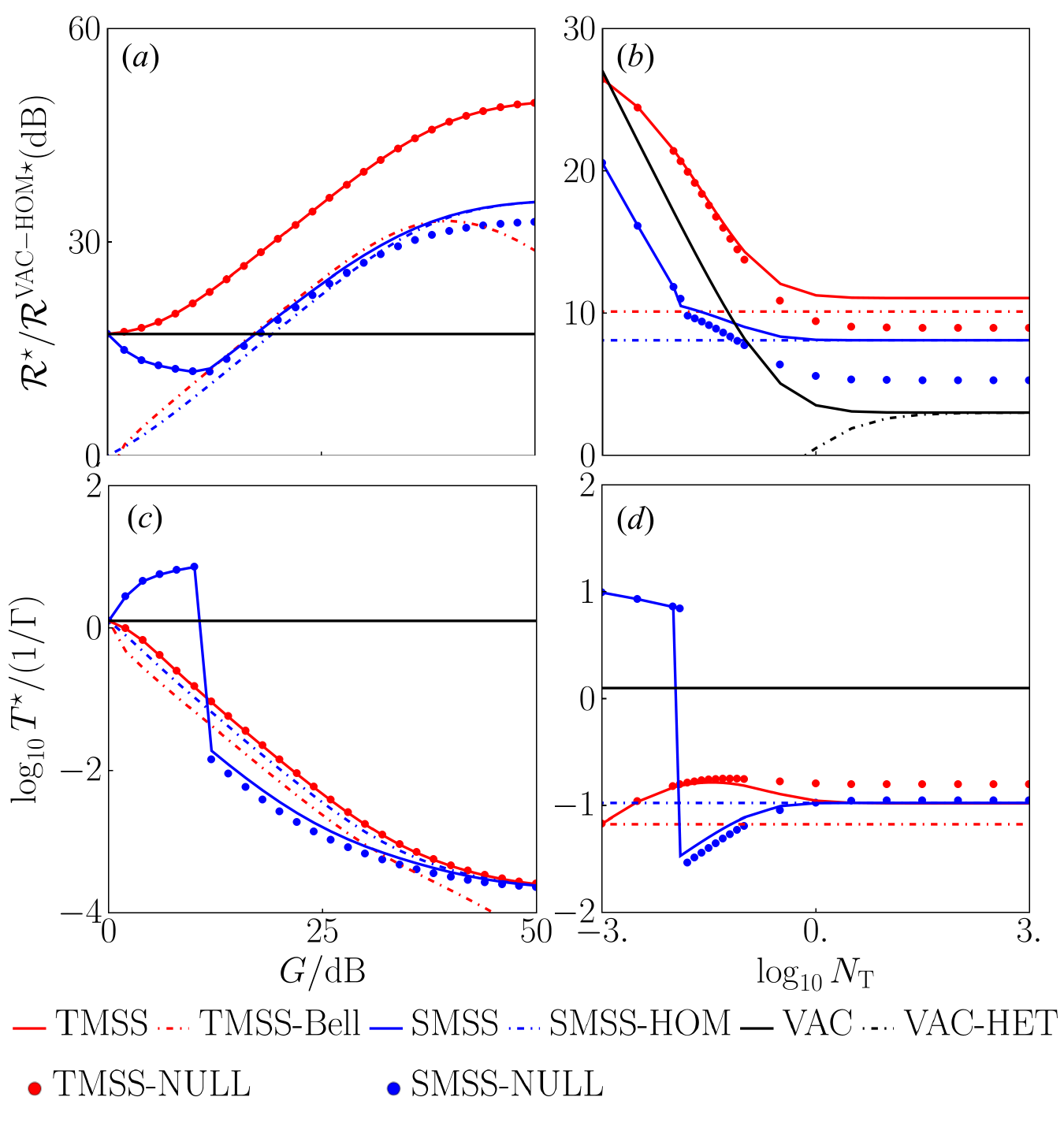}
    \caption{
    (a,b) Quantum advantage in Fisher information rate over the classical protocol using vacuum-state source with homodyne measurement (VAC-HOM), with waiting time $T$ optimized; (c,d) Optimal waiting time $T^\star$. (a,c) Plotted versus squeezing gain $G$ in decibel unit, ${N_{\rm T}}=0.01$; (b,d) Plotted versus noise ${N_{\rm T}}$ in logarithmic scale, $G=10$dB. For all four cases, $\Gamma\tau_A=10^{-4}$, $\Gamma_{\rm idler}\tau_A=10^{-6}$. Solid lines are independent on measurement. Dot-dashed lines are for conventional quantum measurements: SMSS-HOM, SMSS source with homodyne measurement; TMSS-Bell, TMSS source with Bell homodyne measurement~\cite{pirandola2011quantum,shi2020entanglement}; VAC-HET, vacuum source with heterodyne measurement. Dots are for our measurement proposal: TMSS/SMSS-NULL, TMSS/SMSS source with nulling receiver measurement.
    For the nulling receiver, we choose the antisqueezing gain $G^\star$ such that the support of the output state is the photon-number basis, which is asymptotically optimal at the limit ${N_{\rm T}}\to 0$~\cite{gefen2019overcoming}. For large $G$ SMSS-HOM always achieves SMSS-QFI, verified by aymptotic analysis at $G\to \infty$.
    }
    \label{fig:QFI+receiver}
\end{figure}

In Fig.~\ref{fig:QFI+receiver}, we plot the performance of each measurement and the corresponding numerically optimized waiting time. In subplot (a), we consider the low noise limit of ${N_{\rm T}}=0.01$ and plot the QFI rate normalized by the vacuum homodyne QFI rate, versus different squeezing strength $G$. For the vacuum input, as no squeezing is involved, the QFI rate (black solid) has a constant advantage over the vacuum homodyne. At the same time, it is known that the photon counting achieves this QFI rate. As we see in subplots (a) and (b), for SMSS source, when $G$ is small and ${N_{\rm T}}$ is low, the SMSS-NULL detection (blue points) shows advantage over homodyne detection (blue dot-dash). However, SMSS only enjoy advantage over vacuum photon-counting when $G$ is large, where we see homodyne detection provides the same QFI-achieving performance, similar to SMSS-NULL. This can be confirmed by asymptotic analyses in large $G$, and also agrees with the intuition that homodyne is optimal for states with large occupation number (as indicated by Ineq.~\eqref{eq:QReq_SMSSbeatVAC} where $G{N_{\rm T}}\gtrsim 1$). Therefore, homodyne and SMSS~\cite{malnou2019,shi2023ultimate} provides a route to achieve quantum advantage without relying on photon counting in the large gain limit.
On the other hand, the QFI of TMSS (red solid) can only be achieved by the TMSS-NULL (red points), while the homodyne based TMSS-Bell measurement (red dot-dash) shows sub-optimal performance. In fact,TMSS-Bell has similar performance with SMSS.

\section{Broadband performance in scan-rate}
\label{sec:scanrate}

In the section, we compare the in-cavity protocol proposed in this paper, with the input-output protocol analyzed in Ref.~\cite{shi2023ultimate}. In Ref.~\cite{shi2023ultimate}, the probe is first coupled into the cavity, then the in-cavity probe is driven by the axion signal; finally, the probe is coupled out of the cavity for measurement. In the input-output model, one has access to a continuous temporal output, which consists of $T\Delta \omega/2\pi$ modes for observation $T$ within a frequency bin of width $\Delta\omega\gg 1/T$. In contrast, for the in-situ transient model proposed in this paper, only one output, i.e. the total accumulated in-cavity field $A(T)$, is measured after a time $T$. Below we introduce the spectral scan rate as the figure of merit of various dark matter search protocols.

\subsection{Defining the spectral scan rate}
\label{sec:connect2physicalScanrate}

In this section, we begin with a general derivation of the scan-rate, not assuming specific detection approaches; Then, we connect to the in-cavity detection approach.
Consider the axion photons located in a lineshape $N_A^{\omega_{Ac}}(\omega)$ centered at an unknown specific frequency $\omega_{Ac}\in [-B/2,B/2]$. We use the notation $\omega_{Ac}$ to distinguish from the previous notation $\omega_A=\omega_{Ac}-\omega_c$, which is relative to cavity resonance frequency $\omega_c$. Also note that for in-cavity protocol, this is a delta function as the random-phase-induced linewidth has been already included in the calculation of QFI. To scan the spectrum for the axion, the cavity resonance frequency $\omega_c$ is shifted at a constant interval $\epsilon$ after each detection, as a sequence $[-B/2:\epsilon:B/2]$. To cover the whole bandwidth $B$, $B/\epsilon$ rounds of detection is needed. For each detection, we denote the general QFI rate as 
$
{\calJ(\omega_{Ac}-\omega_c,T)}/{T}
$, to distinguish from the in-cavity approach notation of $\calK$. In each round, we assume that one can obtain information for various different frequencies $j\Delta \omega$. Note that in the in-cavity protocol, a delta-function $N_A^{\omega_{Ac}}(\omega)$ dictates that information is only for a single frequency. For $N=B/\epsilon$ rounds, the average QFI rate in the $N$ measurements is 
\bal 
~&\quad\frac{1}{N}\sum_{n=-N/2}^{N/2}  \sum_{j} \frac{\calJ(j\Delta\omega-n\epsilon,T)}{T} |\partial_{N_A}N_A^{\omega_{Ac}}(j\Delta\omega)|^2 \frac{T \Delta\omega}{2\pi} \\
&\simeq  \frac{1}{N\epsilon}\sum_{n=-N/2}^{N/2} \epsilon \int_{-\infty}^{\infty}\frac{d\omega}{2\pi}\calJ(\omega-n\epsilon,T) |\partial_{N_A}N_A^{\omega_{Ac}}(\omega)|^2    \\
&\simeq \frac{1}{B}\int_{-B/2}^{B/2}d\omega_c \int_{-\infty}^{\infty} \frac{d\omega}{2\pi} \calJ(\omega-\omega_c,T) |\partial_{N_A}N_A^{\omega_{Ac}}(\omega)|^2    \\
&= \frac{1}{B} \int_{-\infty}^{\infty}\frac{d\omega}{2\pi}  \left[\int_{-B/2}^{B/2}d\omega_c \calJ(\omega-\omega_c,T) \right]
 |\partial_{N_A}N_A^{\omega_{Ac}}(\omega)|^2   \\
&\simeq \frac{1}{B} \cdot \left[\int_{-\infty}^{\infty}d\omega_c \calJ(\omega_c,T) \right] \cdot \left[\int_{-\infty}^{\infty}\frac{d\omega}{2\pi} |\partial_{N_A}N_A^{\omega_{Ac}}(\omega)|^2   \right] \\
&= \frac{1}{B}\int_{-\infty}^{\infty}d\omega \calJ(\omega,T) \cdot \frac{1}{\tau_A}  \equiv \frac{\bbJ }{B},
\label{eq:scan_def}
\eal
where $\bbJ$ is the spectral scan rate defined in Ref.~\cite{shi2023ultimate}:
\be 
\bbJ\equiv \frac{1}{\tau_A}\int_{-\infty}^\infty \calJ(\omega)d\omega \,,
\label{eq:bbJ}
\ee
which has the unit of Hz/sec, where $\calJ(\omega)=\calJ(\omega,T\to \infty)$. In the first `$\simeq$' of Eq.~\eqref{eq:scan_def}, we have taken the $\Delta\omega\to0$ limit; in the second `$\simeq$', we have taken the $\epsilon\to0$ limit. The last `$\simeq$' is taking the $B\to\infty$ limit. Here we define the coherent time of axion to be $\tau_A=1/\int_{-\infty}^{\infty}\frac{d\omega}{2\pi} |\partial_{N_A}N_A^{\omega_{Ac}}(\omega)|^2$ in a general sense, consistent with the steady-state limit of the phase jump model which yields a Lorenztian lineshape~\cite{dal2021VDM}. Here we have assumed a fixed axion resonant frequency $\omega_{Ac}$ and obtain the scan rate formula. If one assumes a prior on $\omega_{Ac}$ and take a Bayesian approach, the same scan-rate formula can also be obtained due to the frequency translational symmetry of the problem, as we detail in Appendix~\ref{app:Bayesian}.

The per-measurement QFI in Eq.~\eqref{eq:scan_def} general.
In the input-output model of Ref.~\cite{shi2023ultimate}, we have taken the steady-state limit $T\to \infty$, because the per-measurement QFI $\calJ\cdot T\Delta \omega $ is proportional to $T$; while the in-cavity QFI $\calK(\omega,T)$ saturates for large $T$ and we choose $T$ that maximizes the QFI rate $\calK(\omega,T)/T$. Note that, in general one needs frequency-resolved spectral detection to achieve Eq.~\eqref{eq:bbJ}, different from current input-output experiments~\cite{Braggio2024QuAxionSearch}.

For the in-cavity protocol, the first line of Eq.~\eqref{eq:scan_def} (before the continuous limit) can be further simplified by taking a single mode ${T\Delta \omega}/{2\pi}=1$ and $N_A^{\omega_{Ac}}(\omega)=N_A\delta_{\omega,\omega_{Ac}}$ (where $\delta_{x,y}$ is the Kronecker delta with discrete values variables $x,y$), because we have included the lineshape due to phase jump in the calculation of $\calK$. Thus, the average QFI rate for the in-cavity protocol is
\bal 
~&\quad\frac{1}{N}\sum_{n=-N/2}^{N/2}  \frac{\calK(\omega_{Ac}-n\epsilon,T)}{T}\\
&= \frac{1}{N\epsilon}\sum_{n=-N/2}^{N/2} \epsilon \frac{\calK(\omega_{Ac}-n\epsilon,T)}{T}\\
&\simeq \frac{1}{B}\int_{-\infty}^{\infty}d\omega_c \frac{\calK(\omega_{Ac}-\omega_c,T)}{T}\equiv \frac{\bbK }{B},
\eal
where we have defined the spectral scan rate $\bbK $ as the average spectral QFI rate-bandwidth product
\bal 
\bbK &\equiv \int_{-\infty}^{\infty} d\omega \frac{\calK(\omega,T)}{T}
\label{eq:bbK}
\eal
which has the unit Hz/sec. We integrate the QFI rate over the bandwidth, because in the spectral scanning, both the spectral scanning speed and the precision (QFI) of each detection are desired to be maximized. As a figure of merit, the scan rate is as the a precision-bandwidth product per second.

To achieve the same precision requirement in term of square of signal to noise ratio (${\rm SNR}^2$)~\cite{brady2022entangled}, the minimum scanning time is 
\be 
\frac{{\rm SNR}^2\cdot B}{\bbJ}, \frac{{\rm SNR}^2\cdot B}{\bbK}\,,
\ee 
for the input-output protocol and in-cavity protocol respectively.
Now it is clear that $\bbJ,\bbK$ characterizes the rate achieving a precision-bandwidth product. 


\subsection{Bandpass-limited detection versus spectral receivers}

\begin{figure}
    \centering
    \includegraphics[width=0.8\linewidth]{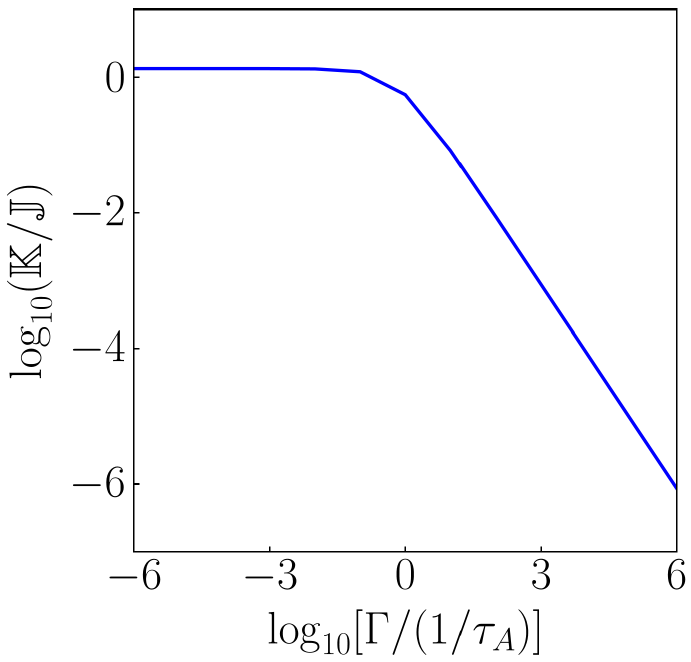}
    \caption{The scan rate ratio of (in-cavity) bandpass-limited photon counting $\bbK$ over (input-output) spectral photon counting $\bbJ$ in logarithmic scale, versus the cavity loss rate $\Gamma$, using a vacuum state input. ${N_{\rm T}}=10^{-4}$. For the input-output method, intrinsic loss rate $\gamma_0=\Gamma$ is chosen to be equal to the in-cavity method, coupling out rate $\gamma_{c}$ is optimized to be $2\gamma_{0}$.}
    \label{fig:Scanrate_IncavOverCoupleout}
\end{figure}

Now we compare a transient in-cavity scan rate versus the input-output scan rate, assuming spectral resolution for the latter. For simplicity, we consider vacuum input, as the relative advantages of SMSS and TMSS over the vacuum limit is understood in previous sections.
Using Eq.(42) in Ref.~\cite{shi2023ultimate}, we obtain the optimal scan rate of the input-output protocol with photon counting on each spectral mode,
\be 
\bbJ=\frac{16 \pi  \gamma _A^2/\tau_A}{27 {N_{\rm T}} \left({N_{\rm T}}+1\right) \left(\gamma _A+\gamma _l\right)}\,.
\ee
In Fig.~\ref{fig:Scanrate_IncavOverCoupleout}, we find that in-cavity detection $\bbK$ and input-output detection $\bbJ$ are comparable at the good cavity limit of $\Gamma\ll 1/\tau_A$. Indeed, $\bbK$ beats $\bbJ$ by a constant factor of $3$dB. 

On the other hand, at the bad cavity limit $\Gamma\gg 1/\tau_A$, spectral photon counting in the input-output protocol ($\bbJ$) beats in-cavity detection ($\bbK$) by a factor of $\Gamma/(1/\tau_A)$. This is because for the input-output protocol, one obtains the output signal over a long integration time window $T\gg 1/\Gamma, 1/\tau_A$, and can presumably count photons in the Fourier domain, which effectively collects all temporal signal bins over time duration of $\tau_A$ coherently. By contrast, the waiting time $T$ in the in-cavity protocol is limited to $1/\Gamma\ll \tau_A$, which effectively operates the parallel bin measurement strategy into $N\simeq \tau_A/(1/\Gamma)$ bins. As predicted by our comparison between the coherent accumulation strategy and the parallel bin measurement strategy (see Appendix~\ref{sec:seqVspar}), coherent collection via spectral photon counting achieves an advantage by a factor $\tau_A\Gamma=\calQ_{\rm axion}/\calQ_{\rm cav}$ when $\calQ_{\rm axion}\gtrsim \calQ_{\rm cav}$. Hence, one desires a spectral photon counter or high-quality signal cavity ($\calQ_{\rm axion}\lesssim \calQ_{\rm cav}$) in order to maximize discovery reach for axion DM searches.

Current axion DM searches (wherein the spectral scan rate is the figure of merit~\cite{chaudhuri2021optimal}) typically operate in the regime $\mathcal{Q}_{\rm cav}/\calQ_{\rm axion}\lesssim 1$~\cite{Lamoreaux2013LinAmpVsPhotonCount}. This constrains discovery potential and, crucially, precludes further enhancements to be gained from photon-counting schemes via quantum probes, in accordance with Ineq.~\eqref{eq:QReq_TMSSbeatVAC_idealidler}. However, this paradigm is beginning to change~\cite{Braggio2024QuAxionSearch}, setting the stage for quantum-enhanced stimulated axion DM searches. 

While an enhancement via quantum probes are achievable for high-quality signal cavities (i.e., $\mathcal{Q}_{\rm cav}/\calQ_{\rm axion}> 1$), there exists a threshold squeezing (or, more generally, a threshold occupation), above which no further enhancement can be gained. This is not an intrinsic limitation but, rather, an artefact of the detection methods currently used (and also considered in this paper). In particular, our detection scheme consists of counting the photons of the cavity mode $\hat{A}(T)$ within a time $T$, thus effectively counting photons within a single bandpass $\sim 1/T$ (cf. Refs.~\cite{Lamoreaux2013LinAmpVsPhotonCount,dixit2021,Braggio2024QuAxionSearch}). Consequentially, the sensitivity of such bandpass-limited measurements depends on the ratio of the axion bandwidth and the signal cavity bandwidth, respectively~\cite{Lamoreaux2013LinAmpVsPhotonCount}. This stimulated bandpass effect, in turn, constrains the advantage to be gained from stimulated DM scanning protocols using bandpass-limited photon counters.

To bypass the bandpass limitation, a spectral photon counter, capable of counting photons in individual frequency bins, is warranted~\cite{Tsang2016SpecAnlyz}. Such a device necessitates, e.g., coupling the signal to an array of narrow-band single photon counters (each centered at different frequencies) or requires an advanced light-matter interface functioning as a multiplexed quantum memory (see, e.g., Section VI.A of Ref.~\cite{gardner2024StochWaveformEst} for further discussion of the latter). A spectral microwave photon counter would be a powerful quantum technology but currently seems out of reach. We have therefore opted to investigate what can be achieved with minimal quantum resources (e.g., bandpass-limited photon counting, as well as Gaussian-state probes), demonstrating regimes of quantum enhancement even under such limitations.\footnote{In contrast, note that linear detection schemes (e.g., HAYSTAC~\cite{backes2021,HAYSTAC2023PhaseII}) measure output fields continuously and, thus, can operate as spectral receivers~\cite{Lamoreaux2013LinAmpVsPhotonCount} (i.e., measure power spectral densities). However, as is well known, such linear detectors are limited by vacuum fluctuations and require large squeezing values to compete with optimal photon counting methods (cf. Ref.~\cite{girvin2016axdm}).}

\section{Discussion and Conclusions}

For entanglement-assisted protocols, the good idler storage cavity may be achieved by superconducting ones since that cavity does not need to be in presence of the magnetic field. In addition, in the distributed sensor network~\cite{brady2022entangled}, where multiple sensors are deployed to search for axion, entanglement ancilla becomes even more practical as a single ancilla storage cavity is sufficient to enhance the entire sensor network's performance. 


Typical stimulated axion DM search experiments, primarily HAYSTAC's squeezed state receiver~\cite{backes2021,HAYSTAC2023PhaseII}, engineer probe states in a different manner than that proposed here. In particular, HAYSTAC's squeezed-state receiver~\cite{backes2021,HAYSTAC2023PhaseII} engineers the quantum state of extra cavity fields---i.e., input and output fields that impinge on and exit from the cavity, respectively---rather than \textit{in situ} engineering of the intra-cavity field considered here (see also Refs.~\cite{wurtz2021cavity,jiang2023}) In principle, there is no difference between extra-cavity state engineering protocols versus \textit{in situ} state engineering protocols. However, in practice, the former may introduce finite ``injection'' losses (e.g., transmission line losses, extra coupling losses etc.) when reflecting itinerant probe states off the cavity. These injection losses will severely limit the performance of Gaussian probe states, in accordance with the Rayleigh curse~\cite{gardner2024StochWaveformEst}. Thus, \textit{in situ} state engineering (analogous to the recent CEASEFIRE protocol~\cite{wurtz2021cavity,jiang2023}) is crucial for Gaussian state probes when photon counting is implemented at the measurement stage.

Finally, we note that there exists other strategies for quantum advantage that can bypass Rayleigh's curse. For instance, a Fock state receiver was recently implemented for a dark photon DM search~\cite{agrawal2024stimulated} (for theoretical analyses, see Refs.~\cite{gorecki2022,gardner2024StochWaveformEst}), which is intrinsically more robust to loss due to the non-Gaussian nature of a Fock state~\cite{gardner2024StochWaveformEst}. With current technology, this requires \QZ{Josephson-junction-based} superconducting elements (e.g., a superconducting control qubit) within the signal cavity to generate the Fock state, thus presenting a hurdle for axion DM searches where the signal cavity is immersed in a strong magnetic field. One could instead prepare propagating microwave photons in a Fock state and count the photons reflected off the signal cavity. In principle, this flying Fock state technique is feasible with an external \QZ{Josephson-junction-based} superconducting device. Though, to the best of our knowledge, such has not been developed that demonstrates both high fidelity and low propagation (or injection) losses for the purpose of quantum sensing.

\begin{acknowledgements}
This material is based upon work supported by the U.S. Department of Energy, Office of Science, National Quantum Information Science Research Centers, Superconducting Quantum Materials and Systems Center (SQMS) under the contract No. DE-AC02-07CH11359. H.S., A.J.B. and Q.Z. also acknowledge the support from National Science Foundation CAREER Award CCF-2240641, National Science Foundation OMA-2326746, 
National Science Foundation Grant No. 2330310, Office of Naval Research Grant
No. N00014-23-1-2296 and Defense Advanced Research Projects Agency (DARPA) under MeasQUIT HR0011-24-9-036. R.D. acknowledges support from the Academy of Finland, grants no. 353832 and 349199. Q.Z. acknowledges discussions with Roni Harnik. 
\end{acknowledgements}

\appendix

\section{Derivation of QFI formulas}
\label{app:derive_QFI}

The full formula for the axion incavity gain is
\begin{widetext}
\bal
~&g(T,\Gamma,\omega_A,\tau_A)\\
&=
\Bigg(4 \Big(\left(\Gamma  \tau _A-2\right) \left(\tau _A^2 \left(4 \omega _A^2+\Gamma ^2\right)+4 \Gamma  \tau
   _A+4\right)+e^{\Gamma  t} \left(\Gamma  \tau _A+2\right) \left(\tau _A^2 \left(4 \omega _A^2+\Gamma ^2\right)-4 \Gamma  \tau
   _A+4\right)\\
   &\quad \quad -2 \Gamma  \tau _A e^{\frac{1}{2} t \left(\Gamma -\frac{2}{\tau _A}\right)} \left(\left(\tau _A^2 \left(4 \omega
   _A^2+\Gamma ^2\right)-4\right) \cos \left(t \omega _A\right)+8 \tau _A \omega _A \sin \left(t \omega
   _A\right)\right)\Big)\Bigg)\\
&\quad ~/\Bigg(\left(e^{\Gamma  t}-1\right) \left(\tau _A^2 \left(4 \omega _A^2+\Gamma ^2\right)-4 \Gamma  \tau
   _A+4\right) \left(\tau _A^2 \left(4 \omega _A^2+\Gamma ^2\right)+4 \Gamma  \tau _A+4\right)\Bigg).
   \label{eq:G_app}
\eal
\end{widetext}

As mentioned in the maintext, we adopt the Gaussian approximation that $a_A$ is in a Gaussian thermal state with mean photon number $N_A g(T,\Gamma,\omega_A,\tau_A)$. Then the in-cavity time evolution Eq.~\eqref{sol} (re-elaborated as below)
\begin{align}
\hat A(T) \simeq &\, \sqrt{\eta(T)} \hat A(0) + \sqrt{1-\eta(T)}\hat a_B (T) \nonumber\\
\quad &+\sqrt{\frac{\gamma_A(1-\eta(T))}{\Gamma}} \hat a_A(T)\,.
\end{align}
represents a bosonic thermal loss channel, with $\hat A(0)$ being the channel input, $\hat a_B(T)$ being the thermal environment, $\sqrt{\frac{\gamma_A(1-\eta(T))}{\Gamma}} \hat a_A(T)$ being a classical additive thermal noise (this is valid because $\gamma_A\ll 1$). 

The calculation of the classical Fisher information for vacuum-state source with homodyne measurement is straightforward. The phase quadrature variance of the output $\hat A(T)$ is
\begin{widetext}
\bal 
\expval{\left(\hat P_A-\expval{\hat P_A}\right)^2}&=
4 \gamma _A N_A \tau _A \Bigg(-\frac{e^{-\frac{1}{2} t \left(\frac{2}{\tau
   _A}+2 i \omega _A+\Gamma \right)}}{\tau _A^2 \left(4 \omega _A^2+\Gamma
   ^2\right)-8 i \tau _A \omega _A-4}-\frac{e^{-\frac{1}{2} t
   \left(\frac{2}{\tau _A}-2 i \omega _A+\Gamma \right)}}{\tau _A^2 \left(4
   \omega _A^2+\Gamma ^2\right)+8 i \tau _A \omega _A-4}\\
&\quad +\frac{e^{\Gamma 
   (-t)} \left(\Gamma  \tau _A-2\right)}{\Gamma  \tau _A \left(\tau _A
   \left(\Gamma  \left(\Gamma  \tau _A-4\right)+4 \tau _A \omega
   _A^2\right)+4\right)}+\frac{\Gamma  \tau _A+2}{\Gamma  \tau _A \left(\tau
   _A \left(\Gamma  \left(\Gamma  \tau _A+4\right)+4 \tau _A \omega
   _A^2\right)+4\right)}\Bigg)+{N_{\rm T}}+\frac{1}{2},
\eal
where $\hat P_A\equiv \sqrt{2}\Im\hat A$. 
The homodyne measurement of $\hat P_A$ yields a random readout $P_A$ subject to the Gaussian distribution $p_{P_A}(x)=\frac{1}{\sqrt{2\pi\sigma^2}}\exp{-\frac{x^2}{2\sigma^2}}$, where $\sigma^2=\expval{\left(\hat P_A-\expval{\hat P_A}\right)^2}$.
Thus the classical Fisher information from the classical vacuum-state homodyne measurement protocol is 
\small
\bal
&\calK_{\rm VAC-HOM}(T,\Gamma,\omega_A,\tau_A)
\equiv \int_{-\infty}^{\infty} dx \left(\frac{d \log p_{P_A}(x)}{dN_A}\right)^2 p_{P_A}(x)\\
&=
\Bigg(32 e^{-2 \Gamma  T} \Big[\Gamma  \tau _A \left(\tau _A^2 \left(\Gamma
   ^2+4 \omega ^2\right)+8 i \omega  \tau _A-4\right) e^{\frac{1}{2} T
   \left(-\frac{2}{\tau _A}+\Gamma -2 i \omega \right)}-e^{\Gamma  T}
   \left(\Gamma  \tau _A+2\right) \left(\tau _A \left(\Gamma  \left(\Gamma 
   \tau _A-4\right)+4 \omega ^2 \tau _A\right)+4\right)\\
   &\qquad +\Gamma  \tau _A
   \left(\Gamma  \tau _A-2 i \omega  \tau _A-2\right) \left(\Gamma  \tau
   _A+2 i \omega  \tau _A+2\right) e^{\frac{1}{2} T \left(-\frac{2}{\tau
   _A}+\Gamma +2 i \omega \right)}-\left(\Gamma  \tau _A-2\right) \left(\tau
   _A \left(\Gamma  \left(\Gamma  \tau _A+4\right)+4 \omega ^2 \tau
   _A\right)+4\right)\Big]^2\Bigg)\\
   &\quad ~
   /\Bigg(\Gamma ^2 \tau _A^2 \left(2
   {N_{\rm T}}+1\right){}^2 \left(\tau _A^4 \left(\Gamma ^2+4 \omega ^2\right)^2-8
   \tau _A^2 \left(\Gamma ^2-4 \omega ^2\right)+16\right){}^2\Bigg)\,.
\eal
\normalsize
Note that in the on-resonance case $\omega_A=0$, the quadrature variance reduces to
\be
\expval{\left(\hat P_A-\expval{\hat P_A}\right)^2}(\omega_A=0)=
\frac{4 \gamma _A N_A e^{-\Gamma  t} \left(-2 \Gamma  \tau _A
   e^{\frac{1}{2} t \left(\Gamma -\frac{2}{\tau _A}\right)}+e^{\Gamma  t}
   \left(\Gamma  \tau _A-2\right)+\Gamma  \tau _A+2\right)}{\Gamma 
   \left(\Gamma ^2 \tau _A^2-4\right)}+{N_{\rm T}}+\frac{1}{2}.
\ee
Then the on-resonance Fisher information reduces to
\be 
\calK_{\rm VAC-HOM}(T,\Gamma,\omega_A=0,\tau_A)=\frac{32 e^{-2 \Gamma  T} \left(-2 \Gamma  \tau _A e^{\frac{1}{2} T
   \left(\Gamma -\frac{2}{\tau _A}\right)}+e^{\Gamma  T} \left(\Gamma  \tau
   _A-2\right)+\Gamma  \tau _A+2\right){}^2}{\Gamma ^2 \tau _A^2 \left(2
   {N_{\rm T}}+1\right){}^2 \left(\Gamma ^2 \tau _A^2-4\right){}^2}.
\ee
\end{widetext}

Now we derive the QFI formulas. Define the operator vector $\bm A \equiv [\hat A, \hat A^\dagger]^T$ ($\bm A\equiv [\hat A, \hat A^\dagger, \hat A_{Anc}, \hat A^\dagger_{Anc}]^T$ for entanglement-assisted case). The Gaussian-state QFI can be easily obtained~\cite{gao2014bounds} given the mean $\bm \mu\equiv \expval{\bm A}$ and covariance matrix $V\equiv \expval{\bm A \bm A^\dagger
}$. 
Over the input mode $\hat A$, the bosonic thermal loss channel fulfills the input-output relation
\be 
\bm x\to \sqrt{\eta(T)} \bm x\,,
\ee
\bal 
~&V\to \\
&\eta(T) V+ [1-\eta(T)]V_B + \frac{\gamma_A(1-\eta(T))}{\Gamma} N_A g(T,\Gamma,\omega_A,\tau_A)\,,
\eal
where $V_B=\bp {N_{\rm T}}+1 &0\\0& {N_{\rm T}}\ep$ is the covariance matrix of the environment,
while over the ancilla mode the channel has no effect (identity channel). Plugging in the input $\bm \mu$ and $V$ of a given source state, we obtain the final output $\bm \mu$ and $V$. Below we show the results for vacuum state, SMSS and TMSS sources.

\begin{widetext}
The vacuum state QFI is
\bal 
\calK_{\rm VAC}(T,\Gamma,\omega_A,\tau_A)=\left[ \frac{\gamma_A(1-\eta(T))}{\Gamma} g(T,\Gamma,\omega_A,\tau_A) \right]^2 \cdot \frac{1}{N_T(1+N_T)},
\eal 
where cavity axion mode gain $g(T,\Gamma,\omega_A,\tau_A)$ is defined in Eq.~\eqref{eq:G_app}.
The SMSS QFI is
\bal 
&\calK_{\rm SMSS}(T,\Gamma,\omega_A,\tau_A)=\\
&\left[ \frac{\gamma_A(1-\eta(T))}{\Gamma} g(T,\Gamma,\omega_A,\tau_A) \right]^2 \cdot \Bigg(2 e^{2 \Gamma  t} \Big(4 G^2 \left(2 N_T \left(N_T+1\right)+1\right) e^{2 \Gamma  t}+(G-1)^2 \left(G^2+1\right) \left(2 N_T+1\right){}^2\\
&\quad +2 (G-1)^2 G \left(2
   N_T+1\right){}^2 e^{\Gamma  t}\Big)\Bigg)\\
&/\Bigg(8 G^2 N_T \left(N_T+1\right) \left(2 N_T \left(N_T+1\right)+1\right) e^{4 \Gamma  t}-2 (G-1)^4 \left(2 N_T+1\right){}^4
   e^{\Gamma  t}\\
&\quad +2 (G-1)^2 G \left(2 N_T+1\right){}^4 e^{3 \Gamma  t}+(G-1)^2 ((G-4) G+1) \left(2 N_T+1\right){}^4 e^{2 \Gamma  t}+(G-1)^4 \left(2 N_T+1\right){}^4\Bigg).
\eal 
The TMSS QFI is
\bal 
&\calK_{\rm TMSS}(T,\Gamma,\omega_A,\tau_A)=\\
&\left[ \frac{\gamma_A(1-\eta(T))}{\Gamma} g(T,\Gamma,\omega_A,\tau_A) \right]^2 \cdot \Bigg( e^{t \left(\Gamma +\Gamma _{\text{idler}}\right)} \Big(4 \left(2 N_T+1\right){}^2 \left(G+e^{\Gamma  t}-1\right) e^{-t \left(\Gamma +3 \Gamma _{\text{idler}}\right)} \left(G+e^{t \Gamma _{\text{idler}}}-1\right) 
\\
&\quad \left(2 (G-1) \left(2 N_T+1\right){}^2 e^{t \Gamma _{\text{idler}}}+(G-1)^2 \left(2 N_T+1\right){}^2+4 N_T \left(N_T+1\right) e^{2 t \Gamma _{\text{idler}}}\right)
\\
&\quad -\!4 \!\left(\!\left(G^2-1\right) \left(2 N_T+1\right)^2 e^{-t \left(\Gamma +\Gamma _{\text{idler}}\right)}\!+\!\left(2
   N_T+1\right)^2 e^{-t \left(\Gamma +3 \Gamma _{\text{idler}}\right)} \left(G\!+\!e^{t \Gamma _{\text{idler}}}\!-\!1\right)^2 \!\left(\left(G^2-1\right) \left(2 N_T+1\right)^2\!\!-\!e^{t \left(\Gamma +\Gamma _{\text{idler}}\right)}\right)\!+\!1\right)\!\Big)\!\Bigg)
\\
&/\Bigg(\left((G-1) \left(2 N_T+1\right)^2 e^{t \Gamma _{\text{idler}}}+(G-1) \left(2 N_T+1\right)^2 e^{\Gamma  t}-2 (G-1) \left(2 N_T+1\right)^2+\left(4 N_T^2+4 N_T+2\right) e^{t \left(\Gamma +\Gamma _{\text{idler}}\right)}\right) 
\\
&\quad \Big(\left(G^2-1\right)^2 \left(2
   N_T+1\right)^4 e^{-2 t \left(\Gamma +\Gamma _{\text{idler}}\right)}-2 \left(G^2-1\right) \left(2 N_T+1\right)^4 \left(G+e^{\Gamma  t}-1\right) e^{-2 t \left(\Gamma +\Gamma _{\text{idler}}\right)} \left(G+e^{t \Gamma _{\text{idler}}}-1\right)
\\
&\quad +2 \left(G^2-1\right) \left(2 N_T+1\right)^2 e^{-t \left(\Gamma +\Gamma _{\text{idler}}\right)}-\left(2 N_T+1\right)^2 e^{-2 t \Gamma _{\text{idler}}} \left(G+e^{t \Gamma _{\text{idler}}}-1\right)^2+\left(2 N_T+1\right)^2 \left(G+e^{\Gamma  t}-1\right)^2 e^{-2 t
   \left(\Gamma +\Gamma _{\text{idler}}\right)} \times 
\\
&\quad \left(2 (G-1) \left(2 N_T+1\right)^2 e^{t \Gamma _{\text{idler}}}+(G-1)^2 \left(2 N_T+1\right)^2+4 N_T \left(N_T+1\right) e^{2 t \Gamma _{\text{idler}}}\right)+1\Big)\Bigg).
\eal

Plugging in Eq.~\eqref{eq:G_app}, we have
\bal 
~&\calK_{\rm VAC}
(T,\Gamma,\omega_A,\tau_A)=\\
&
\Bigg(16 e^{-2 \Gamma  T} \Big(\left(\Gamma  \tau _A-2\right) \left(\tau _A \left(\Gamma  \left(\Gamma 
   \tau _A+4\right)+4 \tau _A \omega_A^2\right)+4\right)-2 \Gamma  \tau _A e^{\frac{1}{2} T \left(\Gamma
   -\frac{2}{\tau _A}\right)} \left(\left(\tau _A^2 \left(\Gamma ^2+4 \omega_A^2\right)-4\right) \cos
   \left(T \omega_A\right)+8 \tau _A \omega_A \sin \left(T \omega_A\right)\right)\\
   &\quad ~~+\left(\Gamma  \tau
   _0+2\right) e^{\Gamma  T} \left(\tau _A \left(\Gamma  \left(\Gamma  \tau _A-4\right)+4 \tau _A \omega
   _0^2\right)+4\right)\Big)^2\Bigg)\\
&/\Bigg(\Gamma ^2 \tau _A^2 {N_{\rm T}} \left({N_{\rm T}}+1\right) \left(\tau _A^4 \left(\Gamma
   ^2+4 \omega_A^2\right)^2-8 \tau _A^2 \left(\Gamma ^2-4 \omega_A^2\right)+16\right)^2\Bigg) \cdot (\gamma_A \tau_A)^2.
\eal
\bal 
~&\calK_{\rm SMSS}(T,\Gamma,\omega_A,\tau_A)=\\
&
\Bigg (64 G^2 \Big(\frac{(2 {N_{\rm T}}+1)^2}{2 G^2}+G (2 {N_{\rm T}}+1)^2 \left(e^{\Gamma
   T}-1\right)+\frac{(2 {N_{\rm T}}+1)^2 \left(e^{\Gamma T}-1\right)}{G}+\frac{1}{2} (2 G
   {N_{\rm T}}+G)^2+\left(4 {N_{\rm T}}^2+4 {N_{\rm T}}+2\right) e^{2 \Gamma T}\\
   &~
   -2 (2 {N_{\rm T}}+1)^2
   e^{\Gamma T}+(2 {N_{\rm T}}+1)^2\Big)
   \Big(\Gamma \tau_A  \left(\Gamma^2
   \tau_A ^2+4 \tau_A ^2 \omega_A^2-8 i \tau_A  \omega_A-4\right) e^{\frac{1}{2} T \left(\Gamma-\frac{2}{\tau_A }+2 i
   \omega_A\right)}\\
   &~+\Gamma \tau_A  \left(\Gamma^2 \tau_A ^2+4 \tau_A ^2 \omega_A^2+8 i \tau_A  \omega_A-4\right)
   e^{\frac{1}{2} T \left(\Gamma-\frac{2}{\tau_A }-2 i \omega_A\right)}\\
   &~-(\Gamma \tau_A +2)
   e^{\Gamma T} \left(\Gamma^2 \tau_A ^2-4 \Gamma \tau_A +4 \tau_A ^2
   \omega_A^2+4\right)-(\Gamma \tau_A -2) \left(\Gamma^2 \tau_A ^2+4 \Gamma \tau_A +4
   \tau_A ^2 \omega_A^2+4\right)\Big)^2\Bigg)\\
   &/\Bigg(\Gamma^2 \tau_A ^2 (-\Gamma \tau_A +2 i \tau_A  \omega_A+2)^2
   (\Gamma \tau_A -2 i \tau_A  \omega_A+2)^2 (\Gamma \tau_A +2 i \tau_A  \omega_A-2)^2 (\Gamma
   \tau_A +2 i \tau_A  \omega_A+2)^2\\
   &\quad\Big(8 G^2 {N_{\rm T}} \left(2 {N_{\rm T}}^3+4 {N_{\rm T}}^2+3 {N_{\rm T}}+1\right) e^{4
   \Gamma T}+(G-1)^2 \left(G^2-4 G+1\right) (2 {N_{\rm T}}+1)^4 e^{2 \Gamma T}-2 (G-1)^4
   (2 {N_{\rm T}}+1)^4 e^{\Gamma T}\\
   &\quad~+2 (G-1)^2 G (2 {N_{\rm T}}+1)^4 e^{3 \Gamma T}+(G-1)^4
   (2 {N_{\rm T}}+1)^4\Big)\Bigg) \cdot (\gamma_A \tau_A)^2.
\eal 

For the TMSS, we present the on-resonance formula $\omega_A=0$ here.
\small
\bal 
&\calK_{\rm TMSS}(T,\Gamma,\Gamma_{\rm idler},\omega_A=0,\tau_A)=\\
&
\Bigg(16 e^{T \left(\Gamma _{\text{idler}}-\Gamma \right)} \left(\Gamma  \tau _A-2 \Gamma  \tau _A e^{\frac{1}{2} T \left(\Gamma
   -\frac{2}{\tau _A}\right)}+e^{\Gamma  T} \left(\Gamma  \tau _A-2\right)+2\right)^2 \Big[4 \left(2 {N_{\rm T}}+1\right)^2
   \left(G+e^{\Gamma  T}-1\right) e^{-T \left(\Gamma +3 \Gamma _{\text{idler}}\right)} \left(G+e^{T \Gamma _{\text{idler}}}-1\right)\\
   &\quad 
   \cdot \left(2 (G-1) \left(2 {N_{\rm T}}+1\right)^2 e^{T \Gamma _{\text{idler}}}+(G-1)^2 \left(2 {N_{\rm T}}+1\right)^2+4 {N_{\rm T}} \left({N_{\rm T}}+1\right) e^{2
   T \Gamma _{\text{idler}}}\right)-4 \Big(\left(G^2-1\right) \left(2 {N_{\rm T}}+1\right)^2 e^{-T \left(\Gamma +\Gamma
   _{\text{idler}}\right)}+\\
   &\quad~ \left(2 {N_{\rm T}}+1\right)^2 e^{-T \left(\Gamma +3 \Gamma _{\text{idler}}\right)} \left(G+e^{T \Gamma
   _{\text{idler}}}-1\right)^2 \left(\left(G^2-1\right) \left(2 {N_{\rm T}}+1\right)^2-e^{T \left(\Gamma +\Gamma_{\text{idler}}\right)}\right)+1\Big)\Big]\Bigg)\\
&/\Bigg(\Gamma ^2 \tau _A^2 \left(\Gamma ^2 \tau _A^2-4\right)^2 \left((G-1) \Big(2
   {N_{\rm T}}+1\right)^2 e^{T \Gamma _{\text{idler}}}+(G-1) \left(2 {N_{\rm T}}+1\right)^2 e^{\Gamma  T}-2 (G-1)
 \left(2
   {N_{\rm T}}+1\right)^2+\left(4 {N_{\rm T}}^2+4 {N_{\rm T}}+2\right) e^{T \left(\Gamma +\Gamma _{\text{idler}}\right)}\Big) \\
&\quad \cdot \Big(\left(G^2-1\right)^2
   \left(2 {N_{\rm T}}+1\right)^4 e^{-2 T \left(\Gamma +\Gamma _{\text{idler}}\right)}-2 \left(G^2-1\right) \left(2 {N_{\rm T}}+1\right)^4
   \left(G+e^{\Gamma  T}-1\right) e^{-2 T \left(\Gamma +\Gamma _{\text{idler}}\right)} \left(G+e^{T \Gamma
   _{\text{idler}}}-1\right)\\
   &\quad \quad +2 \left(G^2-1\right) \left(2 {N_{\rm T}}+1\right)^2 e^{-T \left(\Gamma +\Gamma _{\text{idler}}\right)} -\left(2
   {N_{\rm T}}+1\right)^2 e^{-2 T \Gamma _{\text{idler}}} \left(G+e^{T \Gamma _{\text{idler}}}-1\right)^2\\
   &\quad \quad +\left(2 {N_{\rm T}}+1\right)^2
   \left(G+e^{\Gamma  T}-1\right)^2 e^{-2 T \left(\Gamma +\Gamma _{\text{idler}}\right)} \left(2 (G-1) \left(2 {N_{\rm T}}+1\right)^2 e^{T
   \Gamma _{\text{idler}}}+(G-1)^2 \left(2 {N_{\rm T}}+1\right)^2+4 {N_{\rm T}} \left({N_{\rm T}}+1\right) e^{2 T \Gamma _{\text{idler}}}\right)+1\Big)\Bigg) \\
   &\cdot (\gamma_A \tau_A)^2.
\eal
\normalsize
The TMSS formula for general $\omega_A$ is too lengthy, which is available upon reasonable request to the authors.
\end{widetext}

\section{Sequential coherent accumulation versus parallel bin measurement }
\label{sec:seqVspar}

\begin{figure}
    \centering
    \includegraphics[width=\linewidth]{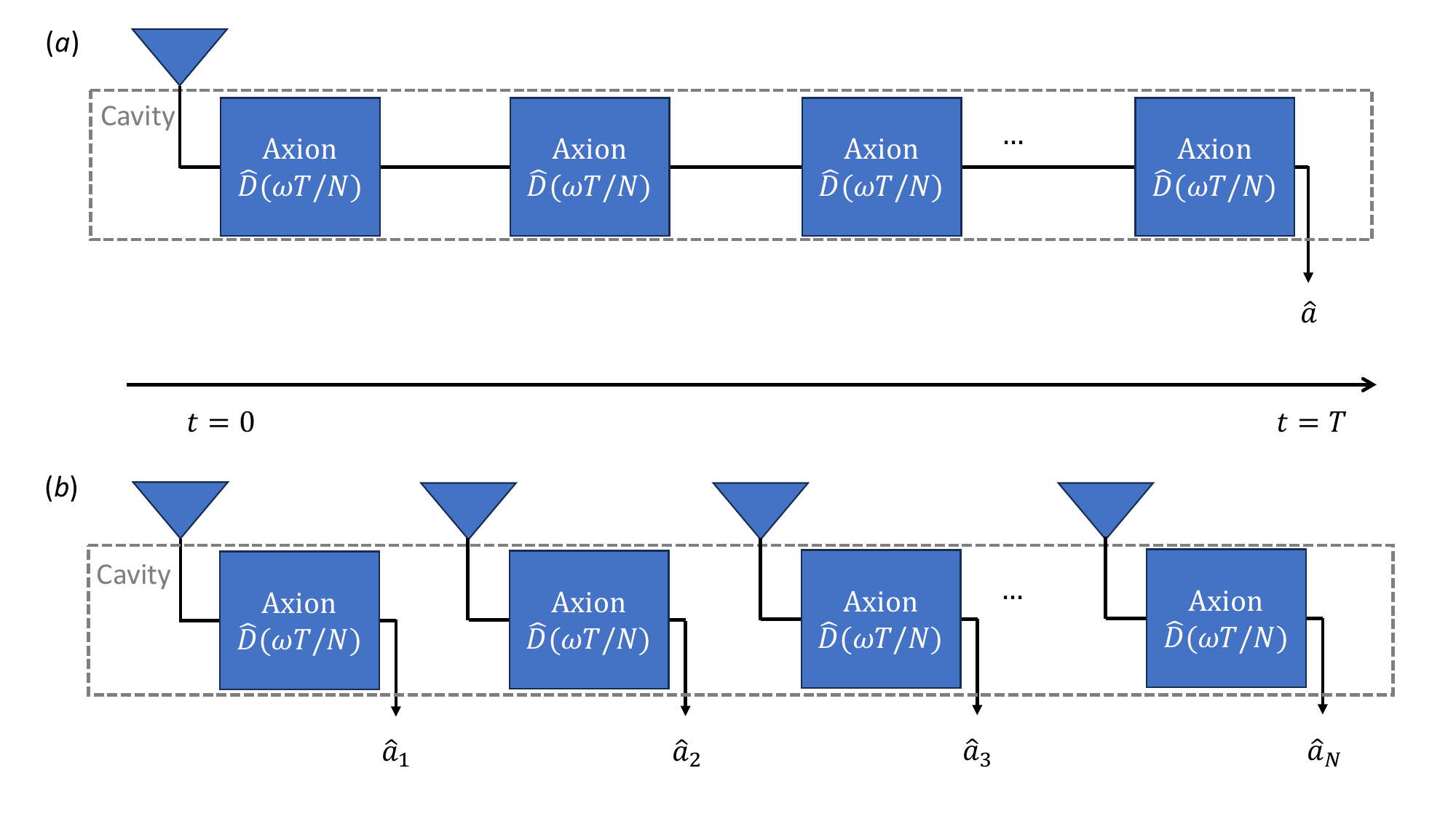}
    \caption{(a) Coherent accumulation strategy: axion displacement is coherently accumulated in the cavity for the whole time duration $T$, which can be alternatively regarded as $N$ sequential displacements over short time bins of duration $T_0\equiv T/N$; (b) parallel bin measurement strategy: after accumulating the displacement for each time bin of duration $T_0$, the in-cavity field is measured, then the accumulation is restarted with a new ideally-stored quantum source. Triangular: Quantum source.
    }    \label{fig:seq_vs_para}
\end{figure}

So far we see that for any cavity loss rate ${\Gamma \tau_A}$, the optimal waiting time $T$ is always around the cavity coherence time $1/\Gamma$. However, existing experiments typically measure for a duration $T\simeq \tau_A$, and $\tau_A\ll 1/\Gamma$ for good cavities.
One may wonder whether it is beneficial to divide the waiting time $T$ into $N$ time bins and measure the $N$-mode state. Specifically, one consider the parallel bin measurement strategy: after accumulating axion signal displacement $\hat D(\Omega T_0)$ for each time bin of duration $T_0\equiv T/N$, where $\Omega$ is the accumulating rate of the axion displacement, one stores the in-cavity signal into a quantum memory ideally loss-free and then restart another round of accumulation. In this case, the quantum source and the axion signal suffers a smaller attenuation $\Gamma T_0$ for a single time bin in this case, c.f. $\Gamma T$ for the worst case in our coherent accumulation protocol. The parallel bin measurement strategy can be regarded as a parallel strategy, while our coherent accumulation strategy can be regarded as a sequential strategy, as shown in Fig.~\ref{fig:seq_vs_para}.

First let us consider classical strategies. It is known that the sequential strategy achieves the Heisenberg limit $1/N$ in error (standard deviation), while the parallel strategy is limited by $1/\sqrt{N}$~\cite{giovannetti2006quantum}. Thus our coherent accumulation strategy, which is sequential, outperforms the parallel bin measurement strategy by $\sqrt{N}$ in precision for large axion coherent time $\tau_A\to \infty$. This can be verified by plugging $T\to N T_0 $ in Eq.~\eqref{eq:nA_eff_lowloss}, the sequential accumulation yields $n_A^{\rm eff}\propto N^2$ (total Fisher $\propto N^4$), while the parallel strategy only achieves $n_A^{\rm eff}\propto N$ (total Fisher $\propto N^3$). For a finite $\tau_A=1$, coherent accumulation never decreases the performance if there is no loss. Indeed, we find that the optimal waiting time that maximizes QFI rate is $T\to\infty$ at the lossless limit $\Gamma T\to 0$. Thus the optimal waiting time is always $T^\star\simeq 1/\Gamma$, beyond which the accumulation of axion signal saturates and the accumulation rate begins to decay.

\section{Discontinuity of optimal waiting time $T$}
\label{app:discontinuity}
Fig.~\ref{fig:advantage_SQZ} in the maintext shows a discontinuity in the optimal waiting time versus varying squeezing gain $G$ using SMSS, similar for Figures~\ref{fig:advantage_TMSS} and \ref{fig:advantage_TMSS_idealidler} using TMSS. Here we provide numerical evidence to explain such discontinuity, for the SMSS as an example. 

\begin{figure}
    \centering \includegraphics[width=.7\linewidth]{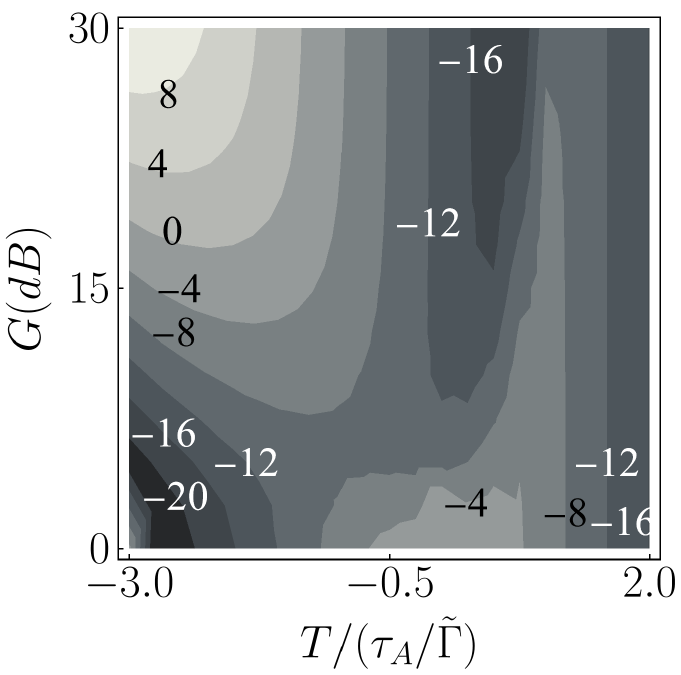}
    \caption{SMSS advantage in dB of QFI rate $\calR$ over vacuum state versus $G$ and waiting time $T$. ${\Gamma \tau_A}=10^{-4}, {N_{\rm T}}=10^{-2}$.   }
    \label{fig:advantage_contour_SQZ}
\end{figure}

As shown in the contour Fig.~\ref{fig:advantage_contour_SQZ}, the SMSS advantage at given $G$ ($\gtrsim 10$dB) have two local optima of $T$, which competes to be the global optima. For smaller $G$, the global optimal $T$ is the larger local optimum; at the discontinuity point, the two local optima gives equal advantage; when $G$ further increases beyond the discontinuity point, the global optimal jumps to the smaller local optimum. 

\section{Justification of Gaussian approximation for non-Gaussian statistics}
\label{sec:justification}

\subsection{Sensing protocol}
\label{sec:protocol}
\begin{figure*}[htbp]
    \centering
    \includegraphics[width=.8\linewidth]{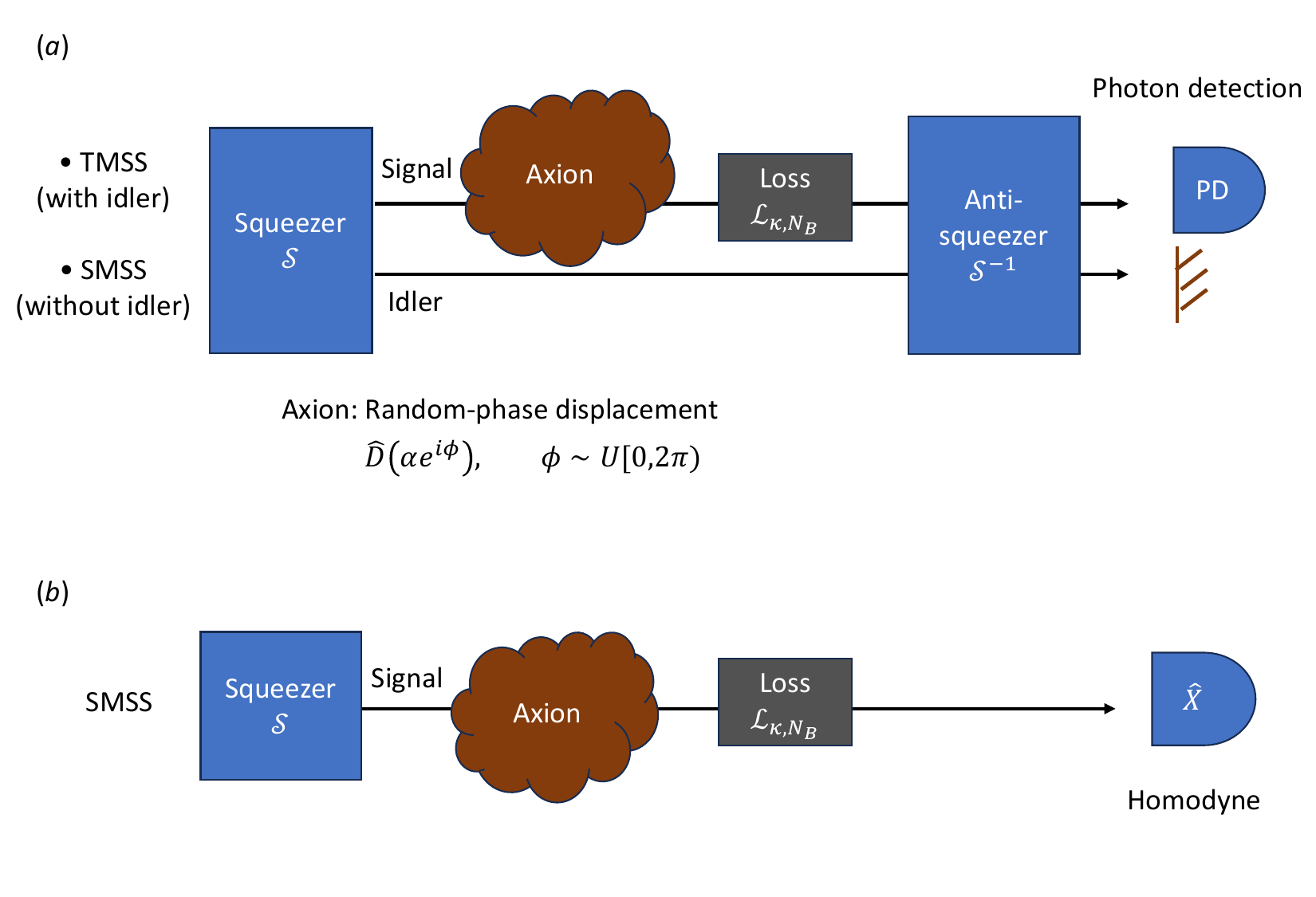}
    \caption{Sensing protocols of the dark matter axion. (a) TMSS/SMSS nulling; (b) SMSS homodyne. TMSS: two-mode squeezed state, SMSS: single-mode squeezed state.}
    \label{fig:protocol}
\end{figure*}

In dark matter search, the axion is coupled to a light probe in cavity and then measured. The interaction is modelled as a displacement channel $\calD_{\alpha e^{i\phi}}$, where $\phi$ is random. Here we assume $\phi$ is uniformly distributed over $[0,2\pi)$, and define $\tilde\calD_{\alpha }(\hat \rho)\equiv \int d\phi \frac{1}{2\pi}\calD_{\alpha e^{i\phi}}(\hat \rho)$. We evaluate the precision of estimating $\alpha$ for various receiver protocols as shown in Fig.~\ref{fig:protocol}. 

In the nulling receiver, a squeezed state of mean photon number $N_S$ is used as the probe signal $\hat a_S$, which can be generated by applying a squeezer $\calS$ to a vacuum mode $\hat v$:
\be 
\hat a_S= \calS (\hat v)\,.
\ee 
Then the probe is displaced by the axion, and suffers loss and additive Gaussian noise modelled by thermal loss channel $\calL_{\kappa,{N_B}}$ where $\kappa$ is transmissivity and ${N_B}$ is the mean photon number of the added noise. The returned mode is 
\be 
\hat a_R=\calL_{\kappa,{N_B}}\circ \tilde \calD_{\alpha }(\hat a_S)\,.
\ee
At the receiver, first the inverse of squeezer $\calS$ is applied to null the returned state back to vacuum
\be
\hat a_R'=\calS^{-1}(\hat a_R)=\calS^{-1}\circ\calL_{\kappa,{N_B}}\circ \tilde \calD_{\alpha }\circ \calS(\hat v)\,.
\ee
If the loss and noise and the axion displacement are negligible, the output is perfectly nulled to vacuum
\be 
\hat a_R'\simeq \calS^{-1}\circ\calS(\hat v)=\hat v\,.
\ee
Finally the photon count of $\hat a_R'$ is measured, which is subject to probability distribution $p(n)$. We denote the nulling receiver using the single-mode squeezed state as SMSS-null.

An improved version of the nulling receiver comes with an idler ancilla, and the single-mode squeezed state is replaced with TMSS. We denote it as TMSS-null.

A well-known sub-optimal sensing protocol utilizes the single-mode squeezed-vacuum, anti-squeezes it at the receiver side, then homodyne it. It turns out that the anti-squeezing does not affect the precision, if other noise sources (e.g. detector noise) not shown in Fig.~\ref{fig:protocol} are negligible. We denote this protocol as SMSS-hom. \\

\begin{figure*}
    \centering
    \includegraphics[width=.8\linewidth]{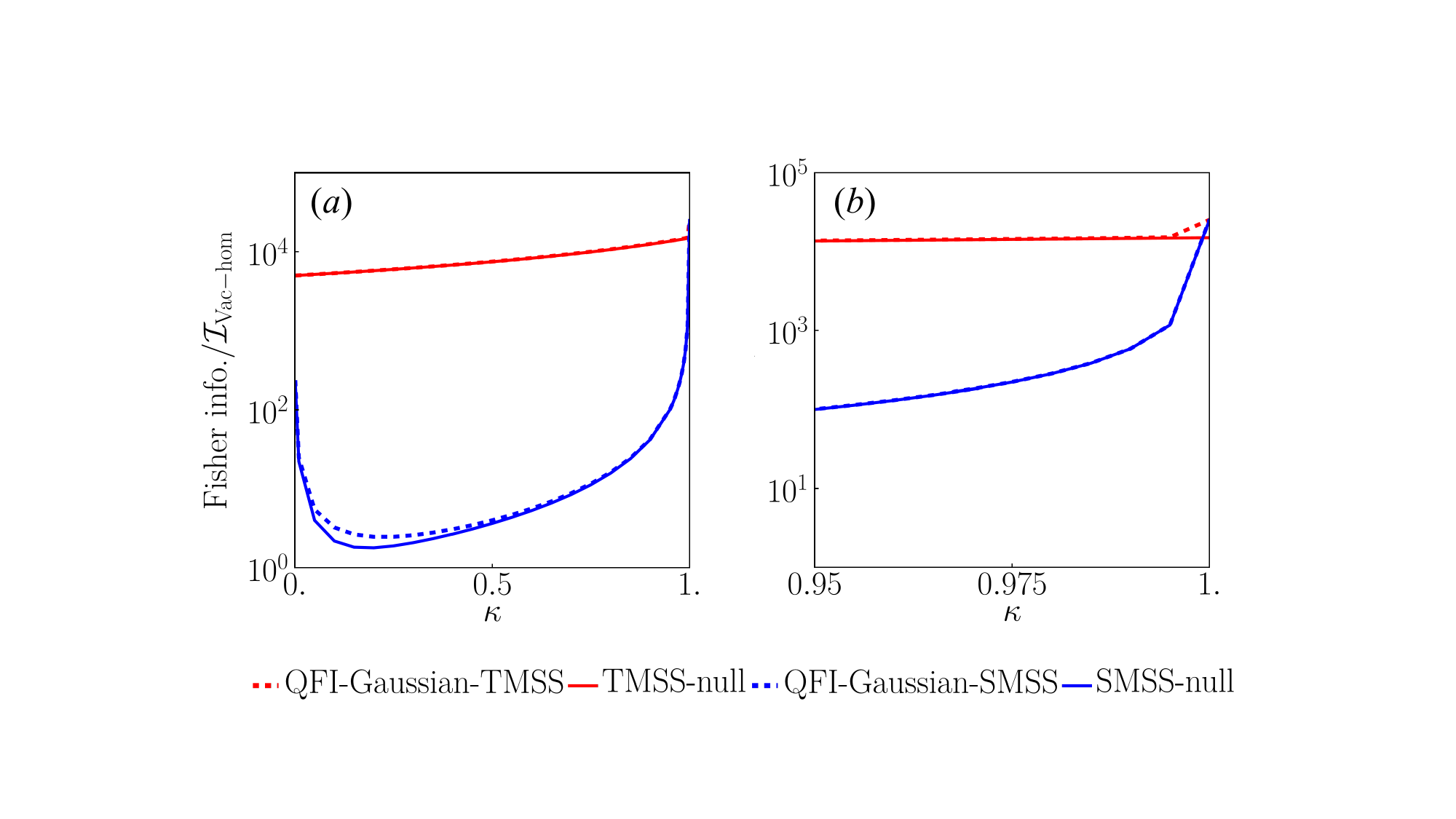}
    \caption{Fisher information of $\alpha$ estimation versus transmissivity $\kappa$, under the random-phase axion model of dark matter search. (a) Overview; (b) Zoom-in at lossless limit $\kappa\to 1$. Normalized by vacuum homodyne Fisher information. TMSS-null, SMSS-null: Nulling receiver using two-mode squeezed state (TMSS) probe and single-mode squeezed state (SMSS) probe. 
    QFI-Gaussian-SMSS/TMSS: the quantum Fisher information of Gaussian axion model, using SMSS/TMSS probe (normalized with vacuum homodyne under the same Gaussian axion model). 
    Thermal noise ${N_B}=0.0001$. Squeezing gain $G=10$. }
    \label{fig:Fisher}
\end{figure*}

The optimal precision of estimating $\alpha$ via the above protocol is the Fisher information
\be 
\calI\equiv \sum_n (\partial_\alpha \log p(n))^2 p(n).
\ee 
We evaluate the Fisher information in Fig.~\ref{fig:Fisher}.

Unfortunately, when using single-mode squeezed-vacuum, the nulled output $\hat a_R'$ carrying the non-Gaussian random-phase axion displacement is in a non-Gaussian state of which the photon count distribution is challenging to derive. Nevertheless, the conditional output $\hat a_{R|\phi}'$ is in a Gaussian state, of which the formulas of conditional photon count distribution $p(n|\phi)$ is much easier to solve. $p(n|\phi)$ is evaluated over the output state conditioned on $\phi$
\be 
\hat a_{R|\phi}'=\calS^{-1}\circ\calL_{\kappa,{N_B}}\circ  \calD_{\alpha e^{i\phi} }\circ \calS(\hat v)\,.
\ee 
and  
\be 
p(n)=\int \frac{d\phi}{2\pi} p(n|\phi)\,.
\ee
In the next section, we review the Gaussian-state results.

\section{Review of Gaussian-state solution}
Ref.~\cite{marian1993squeezed} elaborates the formulas of density matrix in various representations for various Gaussian states subject to the Wigner characteristic function
\be 
\chi(\lambda)=\exp{-(A+\frac{1}{2})|\lambda|^2-\frac{1}{2}[B^*\lambda^2+B(\lambda^*)^2]+C^*\lambda-C\lambda^*}.
\ee
Specifically, Sec.V in Ref.~\cite{marian1993squeezed} derives the photon number distributions. 
The formula is
\be 
p(n)=\sum_{q=0}^n 
\pi  {Q(0)} {\tilde A}^n \frac{1}{q!}  \left(\frac{| {\tilde B}|}{{2\tilde A}}\right)^q  \binom{n}{q} \left| H_q\left(\frac{{\tilde C}}{\sqrt{2} \sqrt{{\tilde B}}}\right)\right|^2,
\ee
where $H_q(\cdot)$ is the $q$th Hermite polynomial,
\begin{align}
&\tilde A=\frac{A (A+1)+| B| ^2}{(A+1)^2+| B| ^2},
\\
&\tilde B= \frac{B}{(A+1)^2+| B| ^2},
\\
&Q(0)=\frac{\exp \left(-\frac{2 (A+1) | C| ^2+C^2 B^*+B \left(C^*\right)^2}{2 \left((A+1)^2-| B| ^2\right)}\right)}{\pi  \sqrt{(A+1)^2-| B| ^2}}.
\end{align}
When the covariance matrix (of any Gaussian state) is given, one immediately obtains the parameters $A,B,C$ and thus the distribution $p(n)$.

For the sensing protocol Sec.~\ref{sec:protocol}, the formula is
\bal
p(n|\phi)=
\sum_{q=0}^n \frac{e^{-\xi } \zeta _1^n \zeta _2^q \binom{n}{q} | H_q(x)| ^2}{ q! \sqrt{d}},
\label{eq:p_cond}
\eal  
where the parameters are defined in Appendix~\ref{app:params}. Here only $\xi$ and $x$ are dependent on $\phi$.


\subsubsection{Parameters for the Gaussian-state photon number distribution}
\label{app:params}
\begin{widetext}
\begin{align}
~&\zeta_1=\frac{2 \kappa  {N_{\rm T}} N_S+{N_{\rm T}}^2+{N_{\rm T}}-(\kappa -1) \kappa  N_S}{2 {N_{\rm T}} \left(\kappa  N_S+N_S+1\right)+{N_{\rm T}}^2+\kappa ^2 \left(-N_S\right)+2 \kappa  N_S^2+2 \kappa  N_S-2 \kappa  \sqrt{N_S^2 \left(N_S+1\right)^2}+N_S+1}
\\
~&\zeta_2=\frac{\left| -\sqrt{N_S \left(N_S+1\right)} \kappa +2 {N_{\rm T}} \sqrt{N_S \left(N_S+1\right)}+\sqrt{N_S \left(N_S+1\right)}\right| }{2 \left(2 \kappa  {N_{\rm T}} N_S+{N_{\rm T}}^2+{N_{\rm T}}-(\kappa -1) \kappa  N_S\right)}
\\
~&d=2 {N_{\rm T}} \left((\kappa +1) N_S+1\right)+{N_{\rm T}}^2+2 \kappa  N_S^2+(1-(\kappa -2) \kappa ) N_S-2 \kappa  \sqrt{N_S^2 \left(N_S+1\right)^2}+1
\\
~&\xi=
\frac{\alpha ^2 \kappa  \left({N_{\rm T}}-\left((\kappa +1) \sqrt{N_S \left(N_S+1\right)} \cos (2 \phi )\right)+\kappa  N_S+N_S+1\right)}{2 {N_{\rm T}} \left((\kappa +1) N_S+1\right)+{N_{\rm T}}^2+\left(-\kappa ^2+2 \kappa +1\right) N_S+2 \kappa  N_S^2-2 \kappa  \sqrt{N_S^2 \left(N_S+1\right)^2}+1} 
\\ 
&x=-\frac{1}{\sqrt{2} \sqrt{N_S \left(N_S+1\right)} \left(-2 {N_{\rm T}}+\kappa -1\right)}\alpha  e^{-i \phi } \\ \nonumber
&\quad \sqrt{\frac{\kappa  \sqrt{N_S \left(N_S+1\right)} \left(2 {N_{\rm T}}-\kappa +1\right)}{2 {N_{\rm T}} \left((\kappa +1) N_S+1\right)+{N_{\rm T}}^2+2 \kappa  N_S^2+(1-(\kappa -2) \kappa ) N_S-2 \kappa  \sqrt{N_S^2 \left(N_S+1\right)^2}+1}} \\ \nonumber
&\quad \Bigg[\sqrt{N_S} \left({N_{\rm T}}+\kappa  N_S\right)+e^{2 i \phi } {N_{\rm T}} \sqrt{N_S+1}+\kappa  e^{2 i \phi } N_S \sqrt{N_S+1}-\kappa  e^{2 i \phi } \sqrt{N_S^2 \left(N_S+1\right)}-\kappa  \sqrt{N_S \left(N_S+1\right)^2}\\ \nonumber 
&\qquad +e^{2 i \phi } \sqrt{N_S+1}\Bigg].
\end{align}
\end{widetext}

\subsection{Evaluating the non-Gaussian photon statistics }
\label{sec:nonGaussian_analytical}
To obtain the unconditional probability from Eq.~\eqref{eq:p_cond} is challenging for large $q$, because it involves an integral of the product of an exponent and high-order Hermite polynomials. 



To tackle this computational complexity, we derive the unconditional probability analytically, which is possible for a monomial (power product) about $\cos\phi$ and $\sin\phi$. This takes a polynomial expansion, which can be done with Taylor expansion of the exponent and binomial expansion. To explore the singularity at $\kappa\to 0$ in Fig.~\ref{fig:Fisher}, we replace the anti-squeezer $\calS^{-1}$ with a tunable squeezer $\calS'(N_{S2})$ parametrized by effective photon number $N_{S2}$. For $\calS'(N_{S2})=\calS^{-1}$, $N_{S2}=N_S$.

Here is a step-by-step guideline of polynomial expansion of Eq.~\eqref{eq:p_cond}. First, the polynomial expansion of $| H_q(x)| ^2$ is
\bal 
&| H_q(x)| ^2=\\
&\sum_{m_1=0}^{\lfloor q/2 \rfloor}\sum_{m_2=0}^{\lfloor q/2 \rfloor}\frac{(-1)^{m_1+m_2} (q!)^2 4^{-m_1-m_2+q} x^{q-2 m_1} \left(x^*\right)^{q-2 m_2}}{m_1! m_2! \left(q-2 m_1\right)! \left(q-2 m_2\right)!},
\eal
which gives two summand indices $m_1,m_2$. Then, the Taylor expansion of $e^{-\xi}$ is
\be 
e^{-\xi}=\sum_{k=0}^{k_c} \frac{1}{k!}(-\xi)^{k},
\ee
where $k_c$ is the cutoff manually chosen to fit the precision requirement. This gives one extra summand index $k$. Note that $\xi$ and $x$ are polynomials of $\cos\phi$, $\sin\phi$, finally a binomial expansion is needed, which gives summand indices $0\le l_1\le q-2m_1, 0\le l_2\le q-2m_2, 0\le l_3\le k$. 

Overall, there are 8 summand indices including $q,n$, the computational complexity is $\sim O(N^8)$, where $N$ is the cutoff for photon count $n$ in Eq.~\eqref{eq:p_cond}.



\section{Bayesian interpretation of the scan rate}
\label{app:Bayesian}

The QFI spectrum is calculated per axion mode, thus to compare with the scan rate Eq.~\eqref{eq:bbK} in this paper, the QFI spectrum is the sum over the measured modes, weighted by the lineshape $|\partial_{N_A}N_A^{\omega_{Ac}}(\omega)|^2$, the squared derivative of the per-mode axion photon spectrum after phase jump 
\be 
N_A^{\omega_{Ac}}(\omega)\to\frac{2N_A}{1+\tau_A^2(\omega-\omega_{Ac})^2}
\ee 
as $T\to \infty$. Here the square is because the Fisher $\calJ_{N_A}(\omega)$ at each frequency mode is proportional to the square of the local signal strength $\calJ_{N_A}(\omega)\propto |\partial_{N_A} N_A^{\omega_{Ac}}(\omega)|^2 \calJ_{N_A^{\omega_{Ac}}}(\omega)$, where $\calJ_{N_A^{\omega_{Ac}}}(\omega)$ is the Fisher estimating the signal $N_A^{\omega_{Ac}}(\omega)$ consistent with $\calJ_{n_a}(\omega)$ in Ref.~\cite{shi2023ultimate}. In Bayesian view, with a uniform prior, the posterior is also uniform of magnitude dependent on the linewidth $2\pi/\tau_A$ of the lineshape, for any $N_A^{\omega_{Ac}}(\omega)$ satisfying $\int_{-\infty}^\infty |\partial_{N_A} N_A^{\omega_{Ac}}(\omega)|^2 d\omega =2\pi/\tau_A$. Below we recover the formula of scan rate in Sec.~\ref{sec:connect2physicalScanrate} in the Bayesian view.



The prior knowledge about the axion center frequency $\omega_{Ac}$ is that it locates in a large bandwidth $B\to\infty$, i.e. $\omega_{Ac}\in [-B/2,B/2]$. As we have almost zero knowledge about the axion, the prior distribution is uniform: $p_{\omega_{Ac}}(\omega)={\rm rect}(\omega/B)/B$. Now consider the measured spectrum, one observes $\frac{T\Delta \omega}{2\pi}$ modes for each measured frequency bin $\Delta \omega\gg 1/T$; thus the per-measurement QFI needs to be summed over the bins, weighted by the squared derivative of axion lineshape $|\partial_{N_A}N_A^{\omega_{Ac}}(\omega)|^2=|\frac{2}{1+\tau_A^2(\omega-\omega_{Ac})^2}|^2$ after involving the phase jump. The Bayesian per-measurement QFI for the steady-state spectral counting model is 
\be 
\int_{-\infty}^\infty d\omega' p_{\omega_{Ac}}(\omega')  \sum_j \calJ(j\Delta\omega) \cdot |\partial_{N_A}N_A^{\omega'}(j\Delta\omega)|^2\cdot \frac{ T\Delta\omega}{2\pi}.
\label{eq:per-meas-QFI}
\ee 
We find it more mathematically convenient to first average the squared derivative spectrum over the prior of the axion center frequency $p_{\omega_{Ac}}(\omega)$, which yields the posterior 
\be 
p_{\rm post}(\omega)=\int_{-\infty}^\infty d\omega' p_{\omega_{Ac}}(\omega') |\partial_{N_A}N_A^{\omega'}(\omega)|^2=2\pi/\tau_AB\,.
\ee 
Here the posterior is always flat, with only the amplitude dependent on the axion signal linewidth $2\pi/\tau_A$. In fact, it is easy to check that this holds for any lineshape $N_A^{\omega_{Ac}}(\omega)$ satisfying $\int_{-\infty}^\infty |\partial_{N_A} N_A^{\omega_{Ac}}(\omega)|^2 d\omega =2\pi/\tau_A$, given uniform prior.
In this case, the Bayesian per-measurement QFI Eq.~\eqref{eq:per-meas-QFI} reduces to simply the sum of per-mode QFIs over the frequency bins weighted by the posterior:
\ba
&&\sum_j p_{\rm post}(j\Delta\omega)\cdot \calJ(j\Delta\omega) \cdot \frac{ T\Delta\omega}{2\pi} 
\nonumber
\\
&&= \frac{ T}{B \tau_A} \int_{-\infty}^{\infty} \calJ(\omega)d\omega, 
\label{eq:totalQFI_coupleout}
\ea
which reduces to an integral for $\Delta\omega\to 0$; finally, multiply Eq.~\eqref{eq:totalQFI_coupleout} by bandwidth coverage $B$ and normalize it by observation time $T\gg 1/\Gamma$, we recover the formula of the scan rate $\bbJ$ Eq.~\eqref{eq:bbJ}. 

On the other hand, for the in-situ 
method proposed in this paper, there is only one accumulated cavity mode to be measured, thus $\frac{T\Delta \omega}{2\pi}=1$. Also, here the phase jump has been included in the derivation of QFI $\calK$ in Sec.~\ref{sec:incavityModel}, thus the input axion photon spectrum is monochromatic: $N_A^{\omega_{Ac},incav}(\omega)=N_A\delta_{\omega,\omega_{Ac}}$. Hence, from Eq.~\eqref{eq:per-meas-QFI}, we obtain the Bayesian per-measurement QFI for the in-cavity protocol
\be 
 \int_{-\infty}^\infty d\omega' p_{\omega_{Ac}}(\omega') \calK(\omega',T) =\frac{1}{B}.
\ee 
Finally, normalizing the per-measurement Fisher information by waiting time $T$ and multiplying it with the bandwidth $B$, we obtain the scan rate Eq.~\eqref{eq:bbK}
\ba
&& \frac{B}{T}\cdot \int_{-\infty}^{\infty} d\omega p_{\omega_{Ac}}(\omega)\calK(\omega,T)
\nonumber
\\
&&\simeq \int_{-\infty}^{\infty} d\omega \frac{\calK(\omega,T)}{T}
\nonumber
\\
&&=\bbK.
\ea


%

\end{document}